% Format AzTeX (F. Arnault)

\catcode`@=11

               %%%%%%%%%%%%%%%%%%%% Fonts %%%%%%%%%%%%%%%%%%%%%

\font\seventeenrm=cmr17
\font\seventeeni=cmmi12 scaled\magstep2
\font\seventeensy=cmsy10  scaled\magstep3
\font\seventeenex=cmex10  scaled\magstep3
\font\seventeengreek=cmr12 scaled\magstep2
\font\seventeenbf=cmbx12 scaled\magstep2
\font\seventeenib=cmmib10 scaled\magstep3
\font\seventeensyb=cmbsy10 scaled\magstep3
\font\seventeenexb=cmexb10 scaled\magstep3

\font\seventeengreekb=cmbx12 scaled\magstep2

\font\fourteenrm=cmr12   scaled\magstep1
\font\fourteeni=cmmi12   scaled\magstep1
\font\fourteensy=cmsy10  scaled\magstep2
\font\fourteenex=cmex10  scaled\magstep2
\font\fourteengreek=cmr12 scaled\magstep1
\font\fourteenbf=cmbx12  scaled\magstep1
\font\fourteensc=cmcsc10 scaled\magstep2
\font\fourteenib=cmmib10 scaled\magstep2

\font\twelverm=cmr12
\font\twelvei=cmmi12
\font\twelvesy=cmsy10  scaled\magstep1
\font\twelveex=cmex10  scaled\magstep1
\font\twelvebf=cmbx12
\font\twelveit=cmti12
\font\twelvebfit=cmbxti12
\font\twelvesl=cmsl12
\font\twelvebfsl=cmbxsl10 scaled\magstep1
\font\twelvett=cmtt12
\font\twelvebftt=cmbtt10 scaled\magstep1
\font\twelvesc=cmcsc10 scaled\magstep1
\font\twelvebfsc=cmbcsc10 scaled\magstep1
\font\twelveib=cmmib10 scaled\magstep1
\font\twelvesyb=cmbsy10 scaled\magstep1
\font\twelveexb=cmexb10 scaled\magstep1
\font\twelvemsam=msam10 scaled\magstep1
\font\twelvebb=msbm10  scaled\magstep1
\font\twelvegoth=eufm10    scaled\magstep1
\font\twelvescript=eusm10  scaled\magstep1
\font\twelvegreek=cmr12
\font\twelvegreekb=cmbx12

\font\tensc=cmcsc10
\font\tenib=cmmib10
\font\tensyb=cmbsy10
\font\tenexb=cmexb10
\font\tenbfsl=cmbxsl10
\font\tenbfit=cmbxti10
\font\tenbfsc=cmbcsc10
\font\tenbftt=cmbtt10
\font\tenmsam=msam10
\font\tenbb=msbm10
\font\tengoth=eufm10
\font\tenscript=eusm10
\font\tengreek=cmr10
\font\tengreekb=cmbx10

\font\ninerm=cmr9
\font\ninei=cmmi9
\font\ninesy=cmsy9

\font\eightrm=cmr8
\font\eighti=cmmi8
\font\eightsy=cmsy8
\font\eightbf=cmbx8
\font\eightsl=cmsl8
\font\eightit=cmti8
\font\eighttt=cmtt8
\font\eightsc=cmcsc8

\font\sevenmsam=msam7
\font\sevenbb=msbm7
\font\sevengoth=eufm7
\font\sevenscript=eusm7
\font\sevengreek=cmr7

\font\fivemsam=msam5

                   % Standard families of fonts (0,1,2,3) %

\def\seventeenm@th{%
 \textfont0=\seventeenrm \scriptfont0=\fourteenrm \scriptscriptfont0=\twelverm
 \textfont1=\seventeeni  \scriptfont1=\fourteeni  \scriptscriptfont1=\twelvei
 \textfont2=\seventeensy \scriptfont2=\fourteensy \scriptscriptfont2=\twelvesy
 \textfont3=\seventeenex \scriptfont3=\fourteenex \scriptscriptfont3=\twelveex}

\def\seventeenm@thbf{%
 \textfont0=\seventeenbf  \scriptfont0=\fourteenbf \scriptscriptfont0=\twelverm
 \textfont1=\seventeenib  \scriptfont1=\fourteenib  \scriptscriptfont1=\twelvei
 \textfont2=\seventeensyb \scriptfont2=\fourteensy \scriptscriptfont2=\twelvesy
 \textfont3=\seventeenexb \scriptfont3=\fourteenex \scriptscriptfont3=\twelveex}

\def\fourteenm@th{%
 \textfont0=\fourteenrm  \scriptfont0=\tenrm  \scriptscriptfont0=\sevenrm
 \textfont1=\fourteeni   \scriptfont1=\teni   \scriptscriptfont1=\seveni
 \textfont2=\fourteensy  \scriptfont2=\tensy  \scriptscriptfont2=\sevensy
 \textfont3=\fourteenex  \scriptfont3=\tenex  \scriptscriptfont3=\tenex}

\def\twelvem@th{%
 \textfont0=\twelverm  \scriptfont0=\ninerm  \scriptscriptfont0=\sevenrm
 \textfont1=\twelvei   \scriptfont1=\ninei   \scriptscriptfont1=\seveni
 \textfont2=\twelvesy  \scriptfont2=\ninesy  \scriptscriptfont2=\sevensy
 \textfont3=\twelveex  \scriptfont3=\tenex   \scriptscriptfont3=\tenex}

\def\twelvem@thbf{%
 \textfont0=\twelvebf  \scriptfont0=\tenbf  \scriptscriptfont0=\sevenbf
 \textfont1=\twelveib  \scriptfont1=\tenib  \scriptscriptfont1=\seveni
 \textfont2=\twelvesyb \scriptfont2=\ninesy  \scriptscriptfont2=\sevensy
 \textfont3=\twelveexb \scriptfont3=\tenex   \scriptscriptfont3=\tenex}

\def\tenm@th{%
 \textfont0=\tenrm  \scriptfont0=\sevenrm  \scriptscriptfont0=\fiverm
 \textfont1=\teni   \scriptfont1=\seveni   \scriptscriptfont1=\fivei
 \textfont2=\tensy  \scriptfont2=\sevensy  \scriptscriptfont2=\fivesy
 \textfont3=\tenex  \scriptfont3=\tenex    \scriptscriptfont3=\tenex}

\def\tenm@thbf{%
 \textfont0=\tenbf  \scriptfont0=\sevenrm  \scriptscriptfont0=\fiverm
 \textfont1=\tenib  \scriptfont1=\seveni   \scriptscriptfont1=\fivei
 \textfont2=\tensyb \scriptfont2=\sevensy  \scriptscriptfont2=\fivesy
 \textfont3=\tenexb \scriptfont3=\tenex    \scriptscriptfont3=\tenex}

\def\eightm@th{%
 \textfont0=\eightrm  \scriptfont0=\fiverm  \scriptscriptfont0=\fiverm
 \textfont1=\eighti   \scriptfont1=\fivei   \scriptscriptfont1=\fivei
 \textfont2=\eightsy  \scriptfont2=\fivesy  \scriptscriptfont2=\fivesy
 \textfont3=\tenex    \scriptfont3=\tenex   \scriptscriptfont3=\tenex}

\def\sevenm@th{%
 \textfont0=\sevenrm  \scriptfont0=\fiverm  \scriptscriptfont0=\fiverm
 \textfont1=\seveni   \scriptfont1=\fivei   \scriptscriptfont1=\fivei
 \textfont2=\sevensy  \scriptfont2=\fivesy  \scriptscriptfont2=\fivesy
 \textfont3=\tenex    \scriptfont3=\tenex   \scriptscriptfont3=\tenex}

%\def\fivem@th{%
% \textfont0=\fiverm   \scriptfont0=\fiverm  \scriptscriptfont0=\fiverm
% \textfont1=\fivei    \scriptfont1=\fivei   \scriptscriptfont1=\fivei
% \textfont2=\fivesy   \scriptfont2=\fivesy  \scriptscriptfont2=\fivesy
% \textfont3=\tenex    \scriptfont3=\tenex   \scriptscriptfont3=\tenex}

                             % font families %

% These families are known by Plain :
% \fam0, \fam1 (\mit, \oldstyle), \fam2 (\cal, symbols), \fam3 (extended sy.)
% \fam\itfam = \fam4 (only \textfont)
% \fam\slfam = \fam5 (only \textfont)
% \fam\bffam = \fam6
% \fam\ttfam = \fam7 (only \textfont)

\newfam\greekf@m
\newfam\msamf@m
\newfam\bbf@m
\newfam\gothf@m
\newfam\scriptf@m

		     % Greek uppercase letters %

\def\Gamma{{\fam\greekf@m\mathchar"7800}}
\def\Delta{{\fam\greekf@m\mathchar"7801}}
\def\Theta{{\fam\greekf@m\mathchar"7802}}
\def\Lambda{{\fam\greekf@m\mathchar"7803}}
\def\Xi{{\fam\greekf@m\mathchar"7804}}
\def\Pi{{\fam\greekf@m\mathchar"7805}}
\def\Sigma{{\fam\greekf@m\mathchar"7806}}
\def\Upsilon{{\fam\greekf@m\mathchar"7807}}
\def\Phi{{\fam\greekf@m\mathchar"7808}}
\def\Psi{{\fam\greekf@m\mathchar"7809}}
\def\Omega{{\fam\greekf@m\mathchar"780A}}

                              % Character sizes %

\def\seventeenpoint{%
  \def\lf{%
     \textfont\greekf@m=\seventeengreek
     \scriptfont\greekf@m=\fourteengreek
     \def\rm{\seventeenm@th\fam0\seventeenrm}%
     \def\it{\sevenit}%
     \def\sl{\sevensl}%
     \def\tt{\seventt}%
     \def\sc{\sevensc}%
     \rm}%
  \def\bf{%
     \textfont\greekf@m=\seventeengreekb
     \scriptfont\greekf@m=\fourteengreek
     %\scriptscriptfont\greekf@m=\twelvegreekb
     \def\rm{\seventeenm@thbf\fam0\seventeenbf}%
     \rm}%
  \normalbaselineskip=20pt\normalbaselines
  \lf}

\def\fourteenpoint{%
  \def\lf{\def\rm{\fourteenm@th\fam0\fourteenrm}\rm}%
  \def\bf{\def\rm{\fourteenbf}\fam0\rm}%
  \def\sc{\fourteensc}%
  \normalbaselineskip=17pt\normalbaselines
  \lf}

\def\twelvepoint{%
  \textfont\msamf@m=\twelvemsam
  \scriptfont\msamf@m=\tenmsam
  %\scriptscriptfont\msamf@m=\sevenmsam
  \textfont\bbf@m=\twelvebb
  \scriptfont\bbf@m=\tenbb
  \scriptscriptfont\bbf@m=\sevenbb
  \textfont\gothf@m=\twelvegoth
  \scriptfont\gothf@m=\tengoth
  \scriptscriptfont\gothf@m=\sevengoth
  \textfont\scriptf@m=\twelvescript
  \scriptfont\scriptf@m=\tenscript
  \scriptscriptfont\scriptf@m=\sevenscript
  \def\lf{%
      \textfont\greekf@m=\twelvegreek
      \scriptfont\greekf@m=\tengreek
      \scriptscriptfont\greekf@m=\sevengreek
      \def\rm{\twelvem@th\fam0\twelverm}%
      \def\it{\twelveit}%
      \def\sl{\twelvesl}%
      \def\tt{\twelvett}%
      \def\sc{\twelvesc}%
      \rm}%
  \def\bf{%
      \textfont\greekf@m=\twelvegreekb
      \scriptfont\greekf@m=\tengreekb
      %\scriptscriptfont\greekf@m=\sevengreekb
      \def\rm{\twelvem@thbf\fam0\twelvebf}%
      \def\it{\twelvebfit}%
      \def\sl{\twelvebfsl}%
      \def\tt{\twelvebftt}%
      \def\sc{\twelvebfsc}%
      \rm}%
  \def\msam{\fam\msamf@m\twelvemsam}%
  \def\bb{\fam\bbf@m\twelvebb}%
  \def\goth{\fam\gothf@m\twelvegoth}%
  \def\script{\fam\scriptf@m\twelvescript}%
  \normalbaselineskip=14pt\normalbaselines
  \lf}

\def\tenpoint{%
  \textfont\msamf@m=\tenmsam
  \scriptfont\msamf@m=\sevenmsam
  \scriptscriptfont\msamf@m=\fivemsam
  \textfont\bbf@m=\tenbb
  \scriptfont\bbf@m=\sevenbb
  %\scriptscriptfont\bbf@m=\fivebb
  \textfont\gothf@m=\tengoth
  \scriptfont\gothf@m=\sevengoth
  %\scriptscriptfont\gothf@m=\fivegoth
  \textfont\scriptf@m=\tenscript
  \scriptfont\scriptf@m=\sevenscript
  %\scriptscriptfont\scriptf@m=\fivescript
  \def\lf{%
      \textfont\greekf@m=\tengreek
      \scriptfont\greekf@m=\sevengreek
      %\scriptscriptfont\greekf@m=\fivegreek
      \def\rm{\tenm@th\fam0\tenrm}%
      \def\it{\tenit}%
      \def\sl{\tensl}%
      \def\tt{\fam\ttfam\tentt}%
      \def\sc{\tensc}%
      \rm}%
  \def\bf{%
      \textfont\greekf@m=\tengreekb
      %\scriptfont\greekf@m=\sevengreekb
      %\scriptscriptfont\greekf@m=\fivegreekb
      \def\rm{\tenm@thbf\fam0\tenbf}%
      \def\it{\tenbfit}%
      \def\sl{\tenbfsl}%
      \def\tt{\fam\ttfam\tenbftt}%
      \def\sc{\tenbfsc}%
      % will redefine \goth,\script ?
      \rm}%
  \def\msam{\fam\msamf@m\tenmsam}%
  \def\bb{\fam\bbf@m\tenbb}%
  \def\goth{\fam\gothf@m\tengoth}%
  \def\script{\fam\scriptf@m\tenscript}%
  \normalbaselineskip=12pt\normalbaselines
  \lf}

\def\eightpoint{%
  \def\lf{\def\rm{\eightm@th\fam0\eightrm}\rm}%
  \def\bf{\def\rm{\eightbf}\rm}%
  \def\it{\eightit}%
  \def\sl{\eightsl}%
  \def\tt{\eighttt}%
  \def\sc{\eightsc}%
  \normalbaselineskip=10pt\normalbaselines
  \setbox\strutbox=\hbox{\vrule height7pt depth3pt width0pt}%
  \lf}

\def\sevenpoint{%
  \def\lf{\def\rm{\sevenm@th\fam0\sevenrm}\rm}%
  \def\bf{\def\rm{\sevenbf}\rm}%
  \def\it{\sevenit}%
  \def\sl{\sevensl}%
  \def\tt{\eighttt}% ersatz
  \normalbaselineskip=8pt\normalbaselines
  \lf}

\tenpoint

\def\mathbf#1{\hbox{\bf$#1$}}

               %%%%%%%%%%%%%%% External files %%%%%%%%%%%%%%%%%

\newif\ifreffile          \reffilefalse
\newread\reffile
\newread\pagenosfile
\newwrite\reffile

\outer\def\openreffile{%
   \re@dreffile
   \immediate\write16{Writing references to \jobname.ref}%
   \immediate\openout\reffile=\jobname.ref
   \reffiletrue}

\def\re@dreffile{%
     \openin\reffile=\jobname.ref
     \ifeof\reffile
        \closein\reffile
        \immediate\write16{*** Not found \jobname.ref ***}%
     \else\closein\reffile
        \immediate\write16{Reading references from \jobname.ref}%
     \let\prefixforref=\relax
     % for some Plain internal cs (dimens,...) ar'nt expanded :
     \catcode`@=11
     \input\jobname.ref
     \catcode`@=12
     \fi}

\newread\xreffile
\outer\def\readref#1{%
   \openin\xreffile=#1.ref
   \ifeof\xreffile
       \closein\xreffile
       \immediate\write16{*** Not found #1.ref ***}%
   \else\closein\xreffile
        \immediate\write16{Reading references from #1.ref}%
        \def\prefixforref{\xchapno.}%
        \catcode`@=11  % for some Plain internal cs (dimens,...) ar'nt expanded
        \input#1.ref
        \catcode`@=12
        \let\prefixforref=\relax
   \fi
   }

               %%%%%%%%%%%%%%%%%% Page setting %%%%%%%%%%%%%%%%

%\hoffset=-3mm    % beurk !
%\voffset=8mm    % (felicie)

%\hsize=6.5in
%\vsize=8.9in
\def\landscape{{\let\tmpsize=\hsize
                \global\let\hsize=\vsize
		\global\let\vsize=\tmpsize}}

\def\t@define{$\spadesuit$}

\def\getp@geno{\global\pageno=\pagenos\chapterno}

\newif\ifshowref    \showreffalse
\let\innerl@bel=\relax
\def\label#1{\def\innerl@bel{#1}\par}  % \par is useful after \subsection
                                       % but I don't now why it works...

                          % headlines and footlines %

\def\frenchmonth{\ifcase\month\or janvier\or f\'evrier\or mars\or avril%
       \or mai\or juin\or juillet\or ao\^ut%
       \or septembre\or octobre\or novembre\or d\'ecembre\fi}
\def\englishmonth{\ifcase\month\or January\or February\or March\or April%
       \or May\or June\or July\or August%
       \or September\or October\or November\or December\fi}
\def\frenchdate{\the\day\ \frenchmonth\ \the\year}
\def\englishdate{\englishmonth\ \the\day, \the\year}
\let\date=\frenchdate
\newif\ifshowdate       \showdatetrue
\def\author{F.~Arnault}
\newif\iffirstpage      \firstpagetrue
\newtoks\firstheadline
\newtoks\rightheadline
\newtoks\leftheadline
\newtoks\firstfootline
\newtoks\otherfootline

\headline={%
   \iffirstpage\the\firstheadline
   \else\ifodd\pageno\the\rightheadline
        \else\the\leftheadline\fi
   \fi}
\firstheadline={\hfil}
\rightheadline={\tenpoint\sl\hfil\firstmark\hfil}
\leftheadline={\tenpoint\sl\hfil\firstmark\hfil}

\footline={%
   \iffirstpage\the\firstfootline\global\firstpagefalse
   \else\the\otherfootline
   \fi}
\firstfootline={\ifshowdate \hfil\sevenrm\author\quad---\quad\date\hfil 
                \else\footfolio \fi}
\otherfootline={\footfolio}
\def\footfolio{\hfil\tenrm --- \folio\ ---\hfil}

			% Chapters, titles, subtitles %

\newskip\titleskipamount
%\titleskipamount=36pt plus4pt minus4pt  % 3\bigskip
\titleskipamount=24pt plus4pt minus4pt  % 2\bigskip
\def\chapterno{\t@define}
\def\ch@pterm@rk{\relax}

\outer\def\chapter#1#2 #3\par{%  %% number, name, title.
   \global\edef\chapterno{#1}%
   \ifnum#1>-1
       \global\edef\ch@pterm@rk{\chapterno.}%
   \fi
   \re@dp@genos
   \getp@geno
   \ifreffile
      \immediate\write\reffile{\def\string\xchapno{#1}}%
      \immediate\write\reffile{\def\string\chap#2{#1}}%
   \fi
   \message{#2 -> chapitre #1 : #3. }%
   % headline of pages 2, 3, ... if no section begins :
   {\def\cr{\relax}\mark{#3}}%
   \leftheadline={\tenpoint\sl\hfil\ch@pterm@rk{\def\cr{\relax} #3}\hfil}
   \ifnum#1>-1
     \leftline{\twelvepoint\bf Chapitre #1}%
     \bigskip
   \fi
   {\twentyonepointb\halign{\centerline{##}\cr#3\crcr}}%
   \ifreffile
      \write\reffile{\string\t@cchapter{\folio}{#1}{#3}}%
   \fi
   \bigskip\bigskip\bigskip
   }

\outer\def\title#1\par{%
   % headline of pages 2, 3, ... if no section begins : 
   {\def\cr{\relax}\mark{#1}}%
   {\seventeenpoint\bf\halign{\centerline{##}\cr#1\crcr}}%
   \ifreffile
      \immediate\write\reffile{\def\string\xchapno{\relax}}%
      \immediate\write\reffile{\string\t@ctitle{#1}}%
   \fi
   \vskip\titleskipamount}

\outer\def\subtitle#1\par{%
   \vskip-\titleskipamount
   {\fourteenpoint\halign{\centerline{##}\cr#1\crcr}}%
   \vskip\titleskipamount}

\parskip=2pt plus 1pt minus 0.5pt   %% Plain : 0pt plus 1pt

				 % Abstracts %

\newdimen\abstractmarginswidth
\abstractmarginswidth=30pt
\long\def\abstract#1.#2\endabstract{%
  \begingroup
  \parindent=\abstractmarginswidth
  \narrower\noindent\eightpoint
  {\bf#1.\enspace}%
  #2\par
  \endgroup}
\def\endabstract{%
 \errhelp={I have encountered a end-of-abstract mark, outside of any abstract}%
 \errmessage{\string\endabstract\space must match an \string\abstract}}

			   % Sections, subsections %

\newif\ifromansection \romansectionfalse
\def\stringsectionnumber#1{\ifromansection\uppercase\expandafter{\romannumeral#1}\else\the#1\fi}

\newcount\sectionnumber
\outer\def\section#1 \par{% % The first <cr> is converted to <sp>_10, knuth p47
    \global\advance\sectionnumber by1
    \global\statnumber=0
    %\bigbreak but always with penalty
    \ifdim\lastskip<\bigskipamount \removelastskip\bigskip \fi
    \penalty-400
    \ifx\innerl@bel\relax\relax
    \else
      \expandafter\xdef\csname sec\innerl@bel\endcsname{\the\sectionnumber}%
      \message{\innerl@bel -> section \the\sectionnumber, }%
      \ifreffile
         \immediate\write\reffile{\edef\string\sec\innerl@bel{%
                          \string\prefixforref\stringsectionnumber{\sectionnumber}}}%
      \fi
    \fi
    \leftline{%
          \ifshowref
            \ifx\innerl@bel\relax\relax  \else\llap{\fiverm\innerl@bel\ }\fi
          \fi
          \twelvepoint\bf\stringsectionnumber{\sectionnumber}. #1}%
    \mark{\ch@pterm@rk\stringsectionnumber{\sectionnumber}. #1}%
    \ifreffile
       \write\reffile{\string\t@csecti@n{\folio}%
                                        \string{#1\string}}%
    \fi
    \nobreak\medskip\noindent
    \let\innerl@bel=\relax
    }

\outer\def\subsection#1 \par{%
    \ifhmode  %% Probably, the \subsection just follows a \section (\noindent)
       \immediate\write-1{%
             I guess that the subsection just follows
             section \the\sectionnumber's title.}%
    \else
        %\bigbreak but always with penalty
        \ifdim\lastskip<\bigskipamount \removelastskip\bigskip \fi
        \penalty-300 % changed from -200
    \fi
    \leftline{\twelvepoint#1}%
    \nobreak\medskip\noindent
    }

\outer\def\subsubsection#1 \par{%
    \ifhmode  %% Probably, the \subsection just follows a \section (\noindent)
       \immediate\write-1{%
             I guess that the subsubsection just follows a section title.}%
    \else
        %\bigbreak but always with penalty
        \ifdim\lastskip<\bigskipamount \removelastskip\bigskip \fi
        \penalty-250 % changed from -200
    \fi
    \leftline{#1}%
    \nobreak\smallskip\noindent
    }

                        % Statements, equations, ... %

\newcount\statnumber

\def\innerstat#1#2{%
    \global\advance\statnumber by1
    \ifx\innerl@bel\relax\relax
    \else
       \expandafter\xdef\csname ref\innerl@bel\endcsname
          {\stringsectionnumber{\sectionnumber}.\the\statnumber}%
       \message{\innerl@bel -> \csname ref\innerl@bel\endcsname, }%
       \ifreffile\immediate\write\reffile{%
             \edef\string\ref\innerl@bel{%
             \string\prefixforref\stringsectionnumber{\sectionnumber}.\the\statnumber}}%
       \fi
    \fi
    \medbreak\noindent
    \ifshowref
       \ifx\innerl@bel\relax\relax \else\llap{\fiverm\innerl@bel\ }\fi
    \fi
    \alphaenumreset
    {\bf\stringsectionnumber\sectionnumber.\the\statnumber. --- #1.\enspace}%
    %{\bf\romansection\the\sectionnumber.\the\statnumber. --- #1.\enspace}%
    {#2\par}%
    \let\innerl@bel=\relax
    \ifdim\lastskip<\medskipamount \removelastskip\penalty55\medskip\fi
    }

\outer\def\stat #1. #2\par{\innerstat{#1}{\sl#2}}
\outer\def\statrm #1. #2\par{\innerstat{#1}{#2}} 
\outer\def\remark #1. #2\par{%
    \smallbreak\noindent
    {\bf#1.\enspace}%
    {#2\par}%
    \ifdim\lastskip<\medskipamount \removelastskip\penalty55\medskip\fi
    }

\newcount\eqnumber
\def\eqdef#1{%
    \global\advance\eqnumber by1
    \expandafter\xdef\csname eq#1\endcsname{\the\eqnumber}%
    \message{#1 -> (\csname eq#1\endcsname), }%
    \ifreffile\immediate\write\reffile{%
          \edef\string\eq#1{\string\prefixforref\the\eqnumber}}%
    \fi
    \eqno\hbox{%
       \lf(\the\eqnumber)%
       \ifshowref\rlap{\fiverm\ #1}\fi}}

% for use in \eqalignno :
\def\eqaligndef#1{% 
    \global\advance\eqnumber by1
    \expandafter\xdef\csname eq#1\endcsname{\the\eqnumber}%
    \message{#1 -> (\csname eq#1\endcsname), }%
    \ifreffile\immediate\write\reffile{\def\string\eq#1{\the\eqnumber}}%
    \fi
    \lf(\the\eqnumber)%
    \ifshowref\rlap{\fiverm\ #1}\fi}

% enumerations
\newcount\enumnumber

\def\enum{%
   \advance\enumnumber by1
   \par\noindent(\the\enumnumber) }

\def\alphaenumreset{\enumnumber="60 }
\def\alphaenum{%
  \advance\enumnumber by1
  \par\noindent(\char\enumnumber) }
\def\rawalphaenum{%
  \advance\enumnumber by1
  (\char\enumnumber) }

			      % Figures %

\newbox\figb@x
\newcount\fignumber
\newdimen\figb@xwidth

\def\figure#1. #2#3{%
  \global\advance\fignumber by1
  \ifx\innerl@bel\relax\relax
  \else
     \expandafter\xdef\csname fig\innerl@bel\endcsname{\the\fignumber}%
     \message{\innerl@bel -> #1 \csname fig\innerl@bel\endcsname, }%
     \ifreffile\immediate\write\reffile{%
         \edef\string\fig\innerl@bel{%
         \string\prefixforref\the\fignumber}}%
     \fi
  \fi
  \setbox\figb@x=\vbox{#3}%
  \figb@xwidth=\wd\figb@x
  \setbox\figb@x=\vbox{%
       \box\figb@x
       \smallskip
       \hbox to\figb@xwidth{%
          \ifshowref
             \ifx\innerl@bel\relax\relax \else\llap{\fiverm\innerl@bel\ }\fi
          \fi
          \hss{\bf #1\ \the\fignumber.} #2\hss}}%
  $$%
    \box\figb@x
  $$%
  \let\innerl@bel=\relax}

                              % Bibliographies %

\newbox\bibb@x
\newcount\bibchosen
\newcount\bibnumber
\outer\def\showbib{\unvbox\bibb@x}

\def\bibdef#1#2#3#4{%
    \global\advance\bibnumber by1
    \expandafter\xdef\csname bib#1\endcsname{\the\bibnumber}%
    \global\setbox\bibb@x=\vbox{%
        \unvbox\bibb@x
        \hang\textindent{[\the\bibnumber]}%
        \strut{\sc#2 :\ \it#3.\/\ \ \rm#4.}%
        \par\medskip}}

\def\bibunlo@ded{\t@define}

\def\bibchoose#1{%
    \global\advance\bibchosen by1
    \expandafter\def\csname bib#1\endcsname{\t@define}}

\let\ifbibchooseall=\iffalse
\def\bibchooseall{\let\ifbibchooseall=\iftrue}

\outer\def\checkbib{\ifbibchooseall\relax
         \else\ifnum\bibchosen=\bibnumber\relax
              \else\immediate\write16{Warning : Some bibliographic 
                                      references remain undefined.}%
              \fi
         \fi}

\outer\def\bibitem#1#2#3#4{%
   \ifbibchooseall\bibdef{#1}{#2}{#3}{#4}%
   \else
       \expandafter\ifx\csname bib#1\endcsname\relax \relax %
       \else
           \expandafter\ifx\csname bib#1\endcsname\bibunlo@ded
               \bibdef{#1}{#2}{#3}{#4}%
           \else\immediate\write16{Warning : attempt to redefine
                                   bibliographic reference #1.}%
           \fi
       \fi
   \fi}

                      % Reference substitution %

% These must be defined AFTER \chapter, \eqdef, \figure...

\def\chapref#1{%
    \expandafter\ifx\csname chap#1\endcsname\relax \t@define
          \immediate\write16{ *** Undefined chapter reference : #1 ***}%
    \else\csname chap#1\endcsname
    \fi}

\def\secref#1{%
    \expandafter\ifx\csname sec#1\endcsname\relax \t@define
          \immediate\write16{ *** Undefined section reference : #1 ***}%
    \else\csname sec#1\endcsname
    \fi}

\def\ref#1{%
    \expandafter\ifx\csname ref#1\endcsname\relax \t@define
          \immediate\write16{ *** Undefined statement reference : #1 ***}%
    \else\csname ref#1\endcsname
    \fi}

\def\eqref#1{%
    \expandafter\ifx\csname eq#1\endcsname\relax \t@define
          \immediate\write16{ *** Undefined equation reference : #1 ***}%
    \else(\csname eq#1\endcsname)%
    \fi}

\def\figref#1{%
    \expandafter\ifx\csname fig#1\endcsname\relax \t@define
          \immediate\write16{ *** Undefined figure reference : #1 ***}%
    \else\csname fig#1\endcsname
    \fi}

\def\bib#1{%
   \expandafter\ifx\csname bib#1\endcsname\relax
      [\t@define]%
      \immediate\write16{ *** Unknown bibliographic reference : #1 ***}%
   \else
      \expandafter\ifx\csname bib#1\endcsname\bibunlo@ded
          \immediate\write16{ *** Unloaded bibliographic reference : #1 ***}%
      \fi
      [\csname bib#1\endcsname]%
   \fi}

               %%%%%%%%%%%%%%% Tables of Contents %%%%%%%%%%%%%

\newif\iftoc   \tocfalse
\newbox\t@cb@x
\newcount\t@csecn@
\newif\ifnarrowtoc  \narrowtoctrue

\def\t@cchapter#1#2#3{%  page, chapno, titre.
  \iftoc{%
     \global\narrowtocfalse
     \def\cr{\relax}%
     \advance\hsize by -\pagecolwidth
     \global\t@csecn@=0
     \parindent=0pt
     \global\setbox\t@cb@x=\vbox{%
         \unvbox\t@cb@x
         \bigskip\bigskip\goodbreak
         \ifnum#2>-1
           \centerline{\bf Chapitre #2}%
           \smallskip\nobreak
         \fi
         \strut\blacksquare\ \bf#3\page{#1}\kern-\pagecolwidth}%
         \smallskip\nobreak}%
  \else\relax
  \fi}

\def\t@ctitle#1{%
  \iftoc{%
     \global\narrowtoctrue
     \parindent=\abstractmarginswidth
     \narrower
     \def\cr{\relax}%
     \advance\hsize by -\pagecolwidth
     \global\t@csecn@=0
     \parindent=0pt
     \global\setbox\t@cb@x=\vbox{%
         \unvbox\t@cb@x
         \medskip
         \strut\blacksquare\ \bf#1}%
         \medskip\nobreak}%
  \else\relax
  \fi}

\def\t@csecti@n#1#2{%
  \iftoc{%
     \advance\hsize by -\pagecolwidth
     \global\advance\t@csecn@ by1
     \parindent=0pt
     \global\setbox\t@cb@x=\vbox{%
         \unvbox\t@cb@x
             \strut
             \ifnarrowtoc\kern\abstractmarginswidth\fi
             \the\t@csecn@. --- #2\page{#1}%
             \ifnarrowtoc\kern\abstractmarginswidth\fi
             \kern-\pagecolwidth}%
          \smallskip}%
  \else\relax
  \fi}

\def\showtoc{%
  \iftoc\unvbox\t@cb@x
  \else
    \immediate\write16{*** TOC is void because you hav'nt said
                       \string\toctrue. ***}%
  \fi}

               %%%%%%%%%%%%%%%%%% \everyjob %%%%%%%%%%%%%%%%%%%

\def\pagenos#1{1\write16{Using default pageno : 1}}

\def\re@dp@genos{%
     \openin\pagenosfile=pagenos.tex
     \ifeof\pagenosfile\closein\pagenosfile
     \else\closein\pagenosfile
        \immediate\write16{Reading pagenos.tex}%
     % pagenos.tex is assumed to contain the following definition
     % \def\pagenos#1{\ifcase#1 n1 \or n2 \or ... \else 1 \fi}%
     \input pagenos.tex
     \fi}

\everyjob={%
     \immediate\write16{Format AzTeX (F. Arnault), version \fmtversion.}
     \setbox\t@cb@x=\vbox{}%
     \setbox\bibb@x=\vbox{}%
     \tracingstats=1}

               %%%%%%%%%%%% Verbatim and Listings %%%%%%%%%%%%%

% see TeXbook, pages 380 -->

% The following cannot be included in a \def because
% the assignment can be parsed only if \obeyspace has expanded.
{\obeyspaces\global\let =\ }  % Active space will have fixed width
% In Plain, active space is \space.  But ' ' is usually not active.
{\def\tmpminus{-\relax}\catcode`-=\active \xdef-{\tmpminus}}
{\def\tmpgreater{>\relax}\catcode`>=\active \xdef>{\tmpgreater}}
{\def\tmplessthan{<\relax}\catcode`<=\active \xdef<{\tmplessthan}}

% Assigns all (special) characters to letter category
\def\unc@tcodespeci@ls{\def\do##1{\catcode`##1=12}\dospecials}

% Verbatim settings will be on before parameter is read
\outer\def\verbatim{\par\begingroup\setupverb@tim\doverbatim}

\def\setupverb@tim{%
  \parindent=0mm
  \def\par{\leavevmode\endgraf}%  % Usual \par is \endgraf
  % Spaces and - will be active :
  \obeylines\unc@tcodespeci@ls\catcode`-=\active\catcode`>=\active
       \catcode`<=\active\obeyspaces\tt}

% The only way to escape is \endverbatim at the end of a line.
% \endverbatim will be seen as a sequence of letters, not a \cs
{\catcode`\|=0 \catcode`\\=12  % Switch escape char to |
 |obeylines|gdef|doverbatim^^M#1\endverbatim
 {|medskip|hrule|nobreak#1|smallskip|nobreak|hrule|medskip|endgroup|noindent}}

\def\endverbatim{%
  \errhelp={I have encountered an end-verbatim mark whereas I was not in
           verbatim mode}%
  \errmessage{\string\endverbatim\space must match a \string\verbati m}}

{\catcode`\|=\active\obeylines%
\gdef\|{\begingroup\catcode`\|=\active\setupverb@tim\let^^M=\ \let|=\endgroup}}

\newcount\l@st@ngl@ne
\outer\def\listing#1#2\par{%
   \par\begingroup\setupverb@tim
   \everypar{\advance\l@st@ngl@ne by1\llap{\sevenrm\the\l@st@ngl@ne\quad}}%
   \medskip
   \line{\leaders\hrule\hfil\quad\sl#1\quad\leaders\hrule\hfil}\nobreak
   \input#2
   \nobreak\smallskip\nobreak\hrule\medskip
   \endgroup}

% Cannot make listing breaking work.
%\newinsert\verbinsert
%\skip\verbinsert=\z@skip
%\count\verbinsert=1000
%\dimen\verbinsert=10pt % only a limit
%\newinsert\lastverbinsert
%\skip\lastverbinsert=\z@skip
%\count\lastverbinsert=1000
%\dimen\lastverbinsert=10pt % only a limit
%\output{\verboutput}
%\def\verboutput{\shipout\vbox{\makeheadline\copy\lastverbinsert\pagebody\copy\verbinsert\makefootline}%
%  \advancepageno
%  \ifnum\outputpenalty>-\@MM \else\dosupereject\fi}

                                 % Hidden parts %

\newif\ifshowhidden  \showhiddenfalse

\def\endhidden{\ifshowhidden\relax\else
  \errhelp={I have encountered a end-of-hidden-part mark, 
            without matching hidden-part mark.}%
  \errmessage{\string\endhidden\space must match a \string\hidde n}\fi}

\outer\def\hidden{\ifshowhidden\relax
                  \else\par\begingroup\s@tuphidd@n\dohidden\fi}

\def\s@tuphidd@n{\obeylines\unc@tcodespeci@ls}  % I don't understand.

{\catcode`\|=0 \catcode`\\=12  % Switch escape char to |
 |obeylines|gdef|dohidden#1\endhidden{|endgroup|fi}}  % why |fi ???
% \endhidden will be seen as a sequence of letters, not a \cs

               %%%%%%%%%%%%%%%%% Initials %%%%%%%%%%%%%%%%%%%%

\font\initialfont=cmbx12 scaled\magstep4
\newbox\initialb@x
\def\initial#1{%
   \setbox\initialb@x=\hbox{\initialfont#1\hskip2pt}%
   \hang\hangafter=-2\hangindent=\wd\initialb@x
   \setbox\initialb@x=\hbox{\hskip-\wd\initialb@x\box\initialb@x}%
   \noindent
   \smash{\lower12pt\box\initialb@x}%
   }
 
\catcode`@=12

               %%%%%%%%%%%%%%%%%%% Divers %%%%%%%%%%%%%%%%%%%%%

\newdimen\pagecolwidth  \pagecolwidth=8mm
\def\page#1{%
       \quad
       %\leaders\hbox to 3mm{\hfil.\hfil}\hfill
       \leaders\hrule height0.1pt\hfill
       %\quad\hbox{#1}}
       \hbox to\pagecolwidth{\hfill#1}}
       
\def\qed{\kern 4pt\penalty500\null\hfill\square\par}

\def\square{%
   \hbox{%
      \vrule
      \vbox to 6pt{\hrule width 4pt\vfill\hrule}%
      \vrule}}

\def\blacksquare{%
  \vbox{%
    \hbox{%
      \kern1pt
      \vrule height5pt depth 0pt width 5pt
      \kern1pt}%
    \kern0.5pt}}

\def\pdem#1-{\vskip-4pt\noindent{\sc#1 }--- \alphaenumreset}
\def\proof{\vskip-3pt\noindent{\sc Proof --- }\alphaenumreset}

\def\bitset{\mathord{\{0,1\}}}
\def\C{{\bb C}}
\def\F{{\bb F}}
\def\N{{\bb N}}
\def\Q{{\bb Q}}
\def\R{{\bb R}}
\def\Z{{\bb Z}}

% See TeXBook p.154.  Category must be 7 to use non-standard \fam (\bb).

  % semi-direct product
  % k with double bar
\def\hbar{{\mathord{\bb\mathchar"717D}}}  % reduced Planck constant (ugly in Plain)
    % 2nd Hebrew letter
   % 3rd Hebrew letter
  % 4th Hebrew letter
                        % \mathchardef\parallel="326B   is in Plain

% \mathchardef\leq="3214 \let\le=\leq     % Plain definitions 
% \mathchardef\geq="3215 \let\ge=\geq
\def\leq{\mathrel{\msam\mathchar"7136}}
\def\geq{\mathrel{\msam\mathchar"713E}}
  % double head arrow

%\def\lshift{\mathop{\mathchar"021C}}   % \ll in plain
%\def\rshift{\mathop{\mathchar"021D}}   % \gg in plain
% see also files msam.map, msbm.map, amsfndoc.dvi

  % sometimes better than |#1|

\def\pairing#1#2{\langle#1,#2\rangle}

\def\assign{\mathrel{:=}}

% \item{#1} defined in plain is also useful !

\def\\{\par}  % hidden \par

% \def\gcd{\mathop{\rm gcd}}  % yet in Plain
    % NOT in Plain

  % \hom is in Plain

% \def\ker{\mathop{\rm ker}\nolimits} % yet in Plain

\def\re{\mathop{\rm Re}}  % j'hesite

\def\fmtversion{2.4.0} % 27/6/11

\def\bra#1{{\langle#1\mathclose\vert}}
\def\ket#1{{\mathopen\vert#1\rangle}}

\title  A complete set of \cr 
        multidimensional Bell inequalities \cr

\vskip-2mm
\centerline{{\twelvepoint Fran\c cois {\sc Arnault}}%
\footnote*{Electronic address: {\tt arnault@unilim.fr}}}
\medskip
\centerline{\eightpoint Universit\'e de Limoges --- XLIM (UMR CNRS 6172)}
\centerline{\eightpoint 123 avenue Albert Thomas, F-87060 Limoges Cedex, France}
\bigskip\bigskip

\bibchoose{ACGKKOZ}
\bibchoose{BBRV}
\bibchoose{Bell}
\bibchoose{Briggs}
\bibchoose{CHSH}
\bibchoose{CGLMP}
\bibchoose{Fine}
\bibchoose{Froissart}
\bibchoose{Fu}
\bibchoose{GargMermin}
\bibchoose{Gottesman}
\bibchoose{Ivanovic}
\bibchoose{KGZMZ}
\bibchoose{KOCCEKO}
\bibchoose{KKCZO}
\bibchoose{Masanes}
\bibchoose{Paterek}
\bibchoose{PitowskySvozil}
\bibchoose{PittengerRubin}
\bibchoose{Peres}
\bibchoose{Schachner}
\bibchoose{Schaefer}
\bibchoose{Schwinger}
\bibchoose{Shchukin}
\bibchoose{SLK06}
\bibchoose{WernerWolf}
\bibchoose{ZukowskiBrukner}
% Bibliographic references for quantum information theory

\bibitem{ACGKKOZ}
 {A.~Ac\'in, J.L.~Chen, N.~Gisin, D.~Kaszlikowski, L.C.~Kwek, C.H.~Oh, M.~\^Zukowski}
 {Coincidence Bell inequality for three three-dimensional systems}
 {Physical Review Letters~92, 250404 (2004)}

\bibitem{ADGL}
 {A.~Ac\'in, T.~Durt, N.~Gisin, J.L.~Latorre}
 {Quantum non-locality in two three level systems}
 {Physical Review A, 65, 052325 (2002)}

\bibitem{BBRV}
 {S.~Bandyopadhyay, P.O.~Boykin, V.~Roychowdhury, F.~Vatan}
 {A new proof for the existence of mutually unbiased bases}
 {Algorithmica~34, 512-528 (2002)}

\bibitem{Bell}
 {J.S.~Bell}
 {On the Einstein Podolsky Rosen paradox}
 {Physics~1, 195 (1964)}

\bibitem{Briggs}
 {W.L.~Briggs}
 {The DFT: an owners' manual for the Discrete Fourier Transform}
 {Society for Industrial Mathematics, 1987}

\bibitem{CHSH}
 {J.F.~Clauser, M.A.~Horne, A.~Shimony, R.A.~Holt}
 {Proposed experiment to test local hidden variables theories}
 {Physical Review Letters~23, 880 (1969)}

\bibitem{CGLMP}
 {D.~Collins, N.~Gisin, N.~Linden, S.~Massar, S.~Popescu}
 {Bell inequalities for arbitrarily high-dimensional systems}
 {Physical Review Letters~88, 040404 (2002)}

\bibitem{Fine}
 {A.~Fine}
 {Hidden variables, joint probabilities, and the Bell inequalities}
 {Physical Review Letters~48, 291 (1982)}

\bibitem{Froissart}
 {M.~Froissart}
 {Constructive Generalization of Bell's inequalities}
 {Nuovo Cimento Della Societ\`a Italiana Di Fisica 64B(2), 241--251 (1981)}

\bibitem{Fu}
 {L-B.~Fu}
 {General correlation functions of the Clauser-Horne-Shimony-Holt inequality for arbitrarily 
  high-dimensional systems}
 {Physical Review Letters~92, 130404 (2004)}

\bibitem{GargMermin}
 {A.~Garg, N.D.~Mermin}
 {Farkas lemma and the nature of reality: statistical implications of quantum correlations}
 {Foundations of Physics 14(1), 1 (1984)}

\bibitem{Gisin}
 {N.~Gisin}
 {Bell inequalities: many questions, a few answers}
 {arXiv:quant-ph/0702021v2 (2008)}

\bibitem{Gottesman}
 {D.~Gottesman}
 {Fault-tolerant quantum computation with higher-dimensional systems}
 {Chaos, Solitons \& Fractals, vol.~10, n.~10, 1749-1758 (1999)}

\bibitem{Ivanovic}
 {I.D.~Ivanovic}
 {Geometrical Description of quantum state determination}
 {Journal of Physics~A, 14, n.~12, 3241--3245}

\bibitem{JLLNL}
 {S-W. Ji, J.~Lee, J.~Lim, K.~Nagata, H-W.~Lee}
 {Multisetting Bell inequality for qudits}
 {Physical Review~A, 78, 052103 (2008)}

\bibitem{KGZMZ}
 {D.~Kaszlikowski, P.~Gnaci\'nski, M.~\^Zukowski, W.~Miklaszewski, A.~Zeilinger}
 {Violations of local realism by two entangled $N$-dimensional systems are stronger than for two
  qubits}
 {Physical Review Letters~85, 4418 (2000)}

\bibitem{KOCCEKO}
 {D.~Kaszlikowski, D.K.L.~Oi, M.~Christandl, K.~Chang, A.~Ekert, L.C.~Kwek, C.H.~Oh}
 {Quantum cryptography based on qutrit Bell inequalities}
 {Physical Review A~67, 012310 (2003)}

\bibitem{KKCZO}
 {D.~Kaszlikowski, L.C.~Kwek, J.L.~Chen, M.~\^Zukowski, C.H.~Oh}
 {Clauser-Horne inequality for three-state systems}
 {Physical Review A~65, 032118 (2002)}

\bibitem{LLD}
 {Y-C.~Liang, C-W.~Lim, D-L.~Deng}
 {Reexamination of a multisetting Bell inequality for qudits}
 {Physical Review~A~80, 052116 (2009)}

\bibitem{Masanes}
 {Ll.~Masanes}
 {Tight Bell inequality for $d$-outcome measurements correlations}
 {Quantum Information \& Computation~3(4), 345 (2003)}

\bibitem{Mermin}
 {N.D.~Mermin}
 {Quantum mechanics vs local realism near the classical limit: a Bell inequality for spin~$s$}
 {Physical Review D~22, 356 (1980)}

\bibitem{Paterek}
 {T.~Paterek}
 {Measurements on composite qudits}
 {Physics Letters~A, 367, 57-64 (2007)}

\bibitem{Peres}
 {A.~Peres}
 {All the Bell inequalities}
 {Foundations of Physics~29, 589-614 (1999)}

\bibitem{PitowskySvozil}
 {I.~Pitowsky, K.~Svozil}
 {Optimal tests for quantum non locality}
 {Physical Review~A~64, 014102 (2001)} 

\bibitem{PittengerRubin}
 {A.O.~Pittenger, M.H.~Rubin}
 {Mutually unbiased bases, generalized spin matrices and separability}
 {Linear Algebra and its Applications~390, 255-278 (2004)}

\bibitem{Ramirez}
 {R.W.~Ramirez}
 {The FFT, Fundamentals and Concepts}
 {Prentice Hall, 1984}

\bibitem{Schachner}
 {G. Schachner}
 {The structure of Bell inequalities}
 {arXiv:quant-ph/0312117 (2003)}

\bibitem{Schaefer}
 {H.H.~Schaefer}
 {Topological Vector Spaces}
 {Springer, Berlin, 1980}

\bibitem{Schwinger}
 {J.~Schwinger}
 {Quantum Mechanics --- Symbolism of Atomic Measurements}
 {edited by B.-G.~Englert, Springer, Berlin, 2001}

\bibitem{Shchukin}
 {E. Shchukin}
 {Bell inequalities, classical cryptography and fractals}
 {arXiv:quant-ph/0703259v2 (2007)}

\bibitem{SLK04}
 {W.~Son, J.~Lee, M.S.~Kim}
 {$d$-outcome measurement for a nonlocality test}
 {Journal of Physics~A: Math. Gen. 37, 11897-11910 (2004)}

\bibitem{SLK06}
 {W.~Son, J.~Lee, M.S.~Kim}
 {Generic Bell inequalities for multipartite arbitrary dimensional systems}
 {Physical Review Letters~96, 060406 (2006)}

\bibitem{WernerWolf}
 {R.F.~Werner, M.M.~Wolf}
 {All-multipartite Bell-correlation inequalities for two dichotomic observables per site}
 {Physical Review A, 64, 032112 (2001)}

\bibitem{ZukowskiBrukner}
 {M.~\^Zukowski, \v C.~Brukner}
 {Bell's theorem for general $N$-qubit states}
 {Physical Review Letters~88, 210401 (2002)}

\checkbib

\def\F{{\script F}}
\def\H{{\script H}}
\def\P{{\script P}}
\def\Q{{\script Q}}
\def\U{{\script U}}
\def\V{{\script V}}
\def\DFT{\mathop{\rm DFT}\nolimits}
\def\Hull{\mathop{\rm Hull}}

\showdatefalse
\openreffile  

\abstract Abstract.  We give a multidimensional generalisation of the complete set of 
Bell-correlation inequalities given by Werner and Wolf in~\bib{WernerWolf}, and by~\^Zukowski and
Brukner in~\bib{ZukowskiBrukner}, for the two-dimensional case.  Our construction applies for the
$n$ parties, two-observables case, where each observable is $d$-valued.  The $d^{d^n}$ inequalities
obtained involve homogeneous polynomials.   They define the facets of a polytope in a complex
vector space of dimension~$d^n$.  We also show that these inequalities are violated by Quantum
Mechanics.  We exhibit examples in the three-dimensional case.

\endabstract

\section Introduction

  The search for Bell inequalities has been the subject of a lot of work.  
Let us recall briefly what the matter is.  Assume that a physical system is made of $n$ subsystems.
For each subsystem, a set of $m$ different observables is considered.  The outcomes of each of the
$nm$ observables belong to a set of cardinality~$d$.  The problem is to find inequalities which
must be satisfied when a local realistic model is assumed.  

  The first such inequalities were provided by 
Bell~\bib{Bell} for the case $(n,m,d)=(2,2,2)$.  It was also shown that Quantum Mechanics 
violate these inequalities.  The CHSH inequalities given in~\bib{CHSH} were shown in~\bib{Fine} to 
be a complete set for the case $(2,2,2)$.  This means that these inequalities provide necessary 
and sufficient conditions for the existence of a local realistic model.

  The authors of~\bib{WernerWolf} and of~\bib{ZukowskiBrukner} gave a complete set of $2^{2^n}$
  Bell inequalities for dichotomic observables, with arbitrary number of parties (case $(n,2,2)$).
The structure of these inequalities was further studied in~\bib{Schachner}, where a recursive 
method to compute Bell inequalities is also given.  The tool for this construction was the 
Walsh-Hadamard transform of Boolean functions.  See also~\bib{Shchukin} which gives some insight 
and useful details.

  A method to obtain a complete set of dichotomic Bell inequalities was given 
in~\bib{PitowskySvozil}.  It has notably been used to exhibit a complete set for the case
$(2,3,2)$.

  The multidimensional case has also been considered in numerous references.  Reasons to explore 
beyond the two-dimensional case include that multidimensional entangled quantum states are known 
to be more resistant to noise, and that they can lead to stronger violations of local 
realism~\bib{KGZMZ}.  Also there are specific uses of the tridimensional case for quantum 
cryptography~\bib{KOCCEKO}.  
The pioneer work for multiple outcome Bell inequalities was~\bib{CGLMP}, where a family of 
multidimensional Bell inequalities, that generalize CHSH, was obtained.  Moreover, these 
inequalities have been later proved tight~\bib{Masanes}.

  However, no complete set has been given yet, beyond the two-dimensional case.

  Instead of the joint probabilities used by many authors for the multi- or three-dimensional case 
(\bib{ACGKKOZ}, \bib{CGLMP}, \bib{KKCZO}, \bib{PitowskySvozil}), we study the correlations between
different observables using correlation functions.  In general, if  $X_i(\lambda)$ and
$Y_j(\lambda)$ are the values obtained by party~$i$ for the observable~$\hat X$ and by party~$j$
for the observable~$\hat Y$, the corresponding correlation is given by
$$
  \int_\Lambda X_i(\lambda) Y_j(\lambda) \rho(\lambda) \,d\lambda
$$
where $\Lambda$ is the domain of the hidden variables~$\lambda$ and~$\rho$ with 
$\int_\Lambda\rho(\lambda)\,d\lambda=1$ is a density function.
These correlation functions have been widely used for the study of the two-dimensional case, where 
the outcomes belong to~$\{\pm1\}$.  We use the same correlation functions also for the 
multidimensional case, but the outcomes are now $d$-th roots of unity in~$\C$.  This approach
has yet been considered for the $d=3$ case in~\bib{Fu}, \bib{KOCCEKO}, \bib{KKCZO}, \bib{SLK06}.  

  We use a geometrical approach. Froissart~\bib{Froissart} has apparently been the first to
do so, and then the authors of~\bib{GargMermin} independently.
It was shown in~\bib{Peres} that the local-realistic domain is a convex polytope (for joint 
probability distributions).  The polytope corresponding to joint probabilities and the one 
corresponding to correlation functions are strongly related because of the relation 
$E(X)=2p(X=1)-1$ between expectation values and probabilities, in the case $d=2$.
The polytope we consider belongs to a complex vector space of dimension~$d^n$. 

  Our inequalities are tight.  This means that they define the facets of the polytope.
The problem of obtaining all the (tight) inequalities was only solved in the 
two-dimensional setting (\bib{Fine}, \bib{PitowskySvozil} with joint probabilities, 
\bib{WernerWolf} and \bib{ZukowskiBrukner} with correlation functions).

  Our inequalities involve products and powers of observables, arranged in homogeneous polynomial 
expressions. Powers of observables have already been used in~\bib{SLK06}.
It turns out that the method developed for $(n,2,2)$ generalizes
pretty well for the multidimensional, two-observables per party case, by means of multidimensional
discrete Fourier transform.  With this tool, we are able to give a complete set of tight Bell
inequalities for the case $(n,2,d)$.

  In this paper, we first presents background about multidimensional Fourier transform (DFT for
short).  Then we recall some facts about the duality of polytopes in (finite dimensional) Hilbert
spaces and study some useful relations between DFT and duality.  Then we produce $d^{d^n}$ Bell
inequalities which generalize those obtained in~\bib{WernerWolf}.  We study the polynomials
involved in these inequalities and give some facts about the symmetries observed.  Then we prove
that our Bell inequalities form a complete set of tight ones.  In section~\secref{violations}, we
explain how violations of our Bell inequalities by Quantum Mechanics can be computed and 
observed.  Finally we explore the case~$d=3$.

\label{MDFT}
\section Multidimensional discrete Fourier transform

  There are numerous references for the discrete Fourier transform.  One of them is~\bib{Briggs}.
However, we give here all the material we need for our purposes.

\subsection Maps from $\Z_d^n$ to the set of $d$-th roots of 1

  The main tool for the classification of dichotomic Bell inequalities is the Walsh-Hadamard
transform for Boolean functions.  For our generalisation of the dichotomic case, we will use
$d$-valued functions and multidimensional discrete Fourier transform.  

  There are two equivalent ways to define Boolean functions: it can be a map $F$ from $\bitset^n$ 
to~$\bitset$ (additive convention), or a map $f$ from~$\bitset^n$ to~$\{1,-1\}$ (multiplicative 
convention).  The equivalence is of course given by~$f=(-1)^F$.  The multiplicative convention is 
more comfortable when dealing with Walsh-Hadamard transforms.  We also adopt a multiplicative
convention, and the considered functions will take their values in the set
$$
  \U = \{1, \omega, \ldots, \omega^{d-1}\}
  \qquad\hbox{where $\omega=\exp(2i\pi/d)$}.
  \eqdef{setU}
$$
We put $\Z_d=\{0,1,\ldots,d-1\}$ and denote by~$\Z_d^n$ the set of $n$-tuples with components 
in~$\Z_d$ ($d,n\in\N^*$).  Also, we denote by~$\F$ or~$\F_{d,n}$ the set of maps from 
$\Z_d^n$ to~$\U$.  There are $d^{d^n}$ such functions.

\subsection The DFT

Let $f$ be a map from $\Z_d^n$ to the complex field $\C$ (or to~$\U$ as a particular case).  The 
(multidimensional) discrete Fourier transform of $f$ is the map~$\DFT f=\hat f$, also from $\Z_d^n$ 
to~$\C$, defined by
$$
  \hat f (r_1,\ldots,r_n)
  =
  \sum_{s_1,\ldots,s_n\in\Z_d} \omega^{r_1s_1+\cdots+r_ns_n} f(s_1,\ldots,s_n)
  \eqdef{DFT}
$$
or, written in compact form, $\hat f(r)=\sum_{s\in\Z_d^n}\omega^{r\cdot s}f(s)$
where $r\cdot s=\sum_{i=1}^n r_is_i$ is the standard scalar product of the $n$-tuples $r$ and $s$.  

  We denote as $H_d$ the matrix $(\omega^{ij})_{0\leq i,j\leq d-1}$.  The $n$-th tensor power of $H_d$
is the $D\times D$ matrix, with $D=d^n$, given by
$$
  H_d^{\otimes n}
  \assign
  (\omega^{r\cdot s})_{r,s\in\Z_d^n}.
$$
The matrices $H_d^{\otimes n}$ can be built up from blocks using recursion on~$n$:
$$
  H_d^{\otimes 0} = (1)
  \qquad\hbox{and}\qquad
  H_d^{\otimes n} = \pmatrix{ \omega^{ij} H_d^{\otimes n-1} }_{0\leq i,j\leq n-1}.
  \eqdef{Hrecursion}
$$
These matrices are a generalization of the usual Hadamard matrices which are obtained in the 
special case $d=2$ (hence $\omega=-1$).

  A map $f$ from $\Z_d^n$ to~$\C$ can be identified to the vector of its values
$(f(s))_{s\in\Z_d^n}$.  The (column) vector of the values of $\hat f$ can be obtained applying
the matrix $H_d^{\otimes n}$ to the (column) vector of the values of $f$~:
$$
  \pmatrix{\hat f(0,0,\ldots,0)\cr \hat f(1,0,\ldots,0)\cr \vdots\cr \hat f(d-1,\ldots,d-1)\cr}
  =
  H_d^{\otimes n}
  \pmatrix{ f(0,0,\ldots,0)\cr f(1,0,\ldots,0)\cr \vdots \cr  f(d-1,\ldots,d-1) \cr}.
$$
Hence, the map~DFT: $f\mapsto\hat f$ is a linear map from~$\C^{d^n}$ to itself.

\subsection Inverse DFT

Let also define ${H_d^*}^{\otimes n}$ the matrix $(\omega^{-r\cdot s})_{r,s\in\Z_d^n}$.  It can be
checked that 
$$
  {H_d^*}^{\otimes n} H_d^{\otimes n} 
  =
  d^nI.
$$
Hence, the inverse transform $\DFT^{-1}$ is obtained by
$$
  f (s_1,\ldots,s_n)
  =
  {1\over d^n} \sum_{r_1,\ldots,r_n\in\Z_d} \omega^{-(r_1s_1+\cdots+r_ns_n)} \hat f(r_1,\ldots,s_n)
$$
or, in compact form, $f(s)={1\over d^n}\sum_{r\in\Z_d^n}\omega^{-r\cdot s}\hat f(r)$.

  In the particular case $d=2$, the multidimensional discrete Fourier transform is the
Walsh-Hadamard transform of Boolean functions:
$$
  \hat f(w) = \sum_{x\in\bitset^n} (-1)^{w\cdot x} f(x)
$$
(using the multiplicative convention: $f(x)\in\{1,-1\}$).

\subsection Some easy results

  Some easy results can be derived from the definition given by Equation~\eqref{DFT}, between the 
discrete Fourier transforms of two elements of~$\F_{d,n}$ which are related in some way:

\label{easyresults}
\stat Proposition.  Put $\hat f=\DFT f$ and~$\hat g=\DFT g$ where $f$ and $g$ belong to $\F_{d,n}$.
\alphaenum If $g(s)=f(-s)$ for all $s\in\Z_d^n$, then $\hat g(r)=\hat f(-r)$ for all $r\in\Z_d^n$.
\alphaenum If $g(s)=f(-s)^*$ for all $s\in\Z_d^n$, then $\hat g(r)=\hat f(r)^*$ for all
$r\in\Z_d^n$ (* denotes complex conjugation).
\alphaenum Let $\delta\in\Z_d^n$.  If $g(s)=f(s+\delta)$ for all $s\in\Z_d^n$ (addition in~$\Z_d^n$ 
is assumed component-wise and modulo~$d$), then $\hat g(r)=\omega^{-r\cdot\delta}\hat f(r)$ for 
all~$r\in\Z_d^n$.
\alphaenum Let $\delta\in\Z_d^n$.  If $g(s)=\omega^{\delta\cdot s}f(s)$ for all $s\in\Z_d^n$, then
$\hat g(r)=\hat f(r+\delta)$ for all $r\in\Z_d^n$.
\alphaenum Let $\sigma$ be a permutation of the set $\{1,\ldots,n\}$.  For 
$s=(s_1,\ldots,s_n)\in\Z_d^n$, we use the shorthand notation
$\sigma(s)=(s_{\sigma(1)},\ldots,s_{\sigma(n)})$.   If $g(s)=f\big(\sigma(s)\big)$ for all
$s\in\Z_d^n$, then $\hat g(r)=\hat f\big(\sigma(r)\big)$ for all $r\in\Z_d^n$.

\proof We show only the last two assertions and leave the first three to the reader.  Assume that
$g(s)=\omega^{\delta\cdot s}f(s)$ for all $s\in\Z_d^n$.  Then
$$
  \hat g(r) 
  = \sum_{s\in\Z_d^n} \omega^{r\cdot s} g(s)
  = \sum_{s\in\Z_d^n} \omega^{r\cdot s} \omega^{\delta\cdot s} f(s)
$$
for all $r\in\Z_d^n$.  Hence,
$$
  \hat g(r - \delta) 
  = \sum_{s\in\Z_d^n} \omega^{(r-\delta)\cdot s} \omega^{\delta\cdot s} f(s)
  = \sum_{s\in\Z_d^n} \omega^{r\cdot s} f(s)
  = \hat f(r).
$$
This proves assertion (d).  Assume now that
$g(s)=f\big(\sigma(s)\big)$ for all $s\in\Z_d^n$.  Then, for all $r\in\Z_d^n$,
$$
  \eqalign{ \hat g(r)
     &= \sum_{s\in\Z_d^n} \omega^{r\cdot s} g(s)
      = \sum_{s\in\Z_d^n} \omega^{r\cdot s} f\big(\sigma(s)\big) \cr
     &= \sum_{s\in\Z_d^n} \omega^{r\cdot\sigma^{-1}(s)} f(s) 
      \qquad\hbox{because $\sigma$ induces a permutation on $\Z_d^n$.}\cr}
$$
Hence
$$
  \hat g\big(\sigma^{-1}(r)\big)
  = \sum_{s\in\Z_d^n} \omega^{\sigma^{-1}(r)\cdot\sigma^{-1}(s)} f(s)
  = \sum_{s\in\Z_d^n} \omega^{r\cdot s} f(s)
  = \hat f(r).
$$
This proves assertion (e).  \qed

\section Convex hulls

  Let $D\in\N$.  We denote $\pairing\beta\gamma=\sum_{i=1}^D\beta^*_i\gamma_i$ the usual
Hermitian inner product in~$\C^D$.  The 
complex vector space~$C^D$ can also be viewed as a vector space over~$\R$, with dimension~$2D$.
Each element $\beta\in\C^D$ can be alternatively written as a $D$-uple $(\beta_1,\ldots,\beta_D)$ of
coordinates belonging to~$\C$ or as a $2D$-uple $(x_1,y_1,\ldots,x_D,y_D)$ of coordinates belonging
to~$\R$, with the relations $\beta_k=x_k+iy_k$.  Recall that the real part of the inner product
$\pairing\cdot\cdot$ is nothing more than the usual scalar product in~$\R^{2D}$:
$$
  \re\pairing\beta\gamma
  =
  \re\sum_{k=1}^D \beta^*_k\gamma_k
  =
  \sum_{k=1}^D (x_k z_k + y_k t_k)
  \qquad\eqalign{
         &\hbox{if $\beta_k=x_k+iy_k$ and $\gamma_k=z_k+it_k$,}\cr
         &\hbox{ with $x_k,y_k,z_k,t_k\in\R$.}\cr}
$$

  Let $S$ be a subset of $\C^D$.  The {\it convex hull} of $S$ is the set
$$
  \Hull S
  \assign
  \bigg\{\sum_k p_k \beta_k 
        \hbox{ with $\beta_k\in S$ and $p_k\in\R_+$ such that }\sum_k p_k=1\bigg\}. 
$$
The {\it dual} (or {\it polar}) of the set~$S$ is, by definition, the set
$$
  T = S^\circ
  \assign
  \{\gamma\in\C^D \mid \re\pairing \beta\gamma \leq 1, \forall \beta\in S\}.
  \eqdef{dualdef}
$$
When $S$ is a polytope containing~0, the vertices of the dual~$T$ correspond to the facets of $S$.
To be precise, $\gamma$ is a vertice of~$T$ if and only if the hyperplane defined by 
the equation $\re\pairing\beta\gamma=1$ contains a facet of~$S$.  

The following result holds (the bipolar Theorem, see~\bib{Schaefer}):

\stat Theorem.  For any subset $S$ of~$\C^D$ containing~0, the dual $S^{\circ\circ}$ of the dual 
of~$S$ is the convex hull of $S$.

\subsection The hull of $\U$ and its dual

We assume here $d>2$.
The convex hull of the set $\U$ is a regular polygon.  The dual of $\U$ is also a
regular polygon with $d$ vertices (see Figure~\figref{polygons}):

\label{Ucirc}
\stat Lemma.  The dual~$\U^\circ$ of $\U$ (with $d>2$) is the polygon with vertices set:
$$
  \V =  \Bigg\{ {1\over\cos(\pi/d)} \exp\bigg({2k+1\over d}i\pi\bigg)
  \quad\Bigg|\quad  k=0,\ldots,d-1 \Bigg\}.
$$

\proof For $\beta_k=\exp\Big({2ki\pi\over d}\Big)\in\U$ and 
$\gamma_l=\exp\Big({(2l+1)i\pi\over d}\Big)\Big/\cos\Big({\pi\over d}\Big)\in\V$ we have
$$
  \re\pairing{\beta_k}{\gamma_l}
  =
  {\re\big( \exp(-2ki\pi/d) \exp\big((2l+1)i\pi/d\big)\big)  \over \cos(\pi/d)}
  =
  {\cos\big((2l+1-2k)\pi/d\big) \over \cos(\pi/d)}.
$$
Thus, $\re\pairing{\beta_k}{\gamma_l}=1$ when $k=l$ or when $k=l+1$ (the vertice~$\gamma_l$
of~$\U^\circ$ corresponds to the edge $\delta_l=(\beta_l$, $\beta_{l+1})$ of~$\Hull\U$).  For the 
other values of $k$, we have $\re\pairing{\beta_k}{\gamma_l}<1$ because $\beta_k$ is in the 
half-plane delimited by $\delta_l$ and containing~0.  \qed
\midinsert
\label{polygons}
\expandafter\ifx\csname graph\endcsname\relax
   \csname newbox\expandafter\endcsname\csname graph\endcsname
\fi
\ifx\graphtemp\undefined
  \csname newdimen\endcsname\graphtemp
\fi
\expandafter\setbox\csname graph\endcsname
 =\vtop{\vskip 0pt\hbox{%
    \special{pn 8}%
    \special{ar 750 900 10 10 0 6.28319}%
    \special{sh 1.000}%
    \special{pn 1}%
    \special{pa 1400 875}%
    \special{pa 1500 900}%
    \special{pa 1400 925}%
    \special{pa 1400 875}%
    \special{fp}%
    \special{pn 8}%
    \special{pa 0 900}%
    \special{pa 1400 900}%
    \special{dt 0.050}%
    \special{sh 1.000}%
    \special{pn 1}%
    \special{pa 725 250}%
    \special{pa 750 150}%
    \special{pa 775 250}%
    \special{pa 725 250}%
    \special{fp}%
    \special{pn 8}%
    \special{pa 750 1650}%
    \special{pa 750 250}%
    \special{dt 0.050}%
    \special{pa 1250 900}%
    \special{pa 750 400}%
    \special{fp}%
    \special{pa 750 400}%
    \special{pa 250 900}%
    \special{fp}%
    \special{pa 250 900}%
    \special{pa 750 1400}%
    \special{fp}%
    \special{pa 750 1400}%
    \special{pa 1250 900}%
    \special{fp}%
    \special{pa 1250 400}%
    \special{pa 250 400}%
    \special{da 0.050}%
    \special{pa 250 400}%
    \special{pa 250 1400}%
    \special{da 0.050}%
    \special{pa 250 1400}%
    \special{pa 1250 1400}%
    \special{da 0.050}%
    \special{pa 1250 1400}%
    \special{pa 1250 400}%
    \special{da 0.050}%
    \special{ar 3750 900 10 10 0 6.28319}%
    \special{sh 1.000}%
    \special{pn 1}%
    \special{pa 4400 875}%
    \special{pa 4500 900}%
    \special{pa 4400 925}%
    \special{pa 4400 875}%
    \special{fp}%
    \special{pn 8}%
    \special{pa 3000 900}%
    \special{pa 4400 900}%
    \special{dt 0.050}%
    \special{sh 1.000}%
    \special{pn 1}%
    \special{pa 3725 250}%
    \special{pa 3750 150}%
    \special{pa 3775 250}%
    \special{pa 3725 250}%
    \special{fp}%
    \special{pn 8}%
    \special{pa 3750 1650}%
    \special{pa 3750 250}%
    \special{dt 0.050}%
    \special{pa 4250 900}%
    \special{pa 3905 424}%
    \special{fp}%
    \special{pa 3905 424}%
    \special{pa 3345 606}%
    \special{fp}%
    \special{pa 3345 606}%
    \special{pa 3345 1194}%
    \special{fp}%
    \special{pa 3345 1194}%
    \special{pa 3905 1376}%
    \special{fp}%
    \special{pa 3905 1376}%
    \special{pa 4250 900}%
    \special{fp}%
    \special{pa 4250 537}%
    \special{pa 3559 312}%
    \special{da 0.050}%
    \special{pa 3559 312}%
    \special{pa 3132 900}%
    \special{da 0.050}%
    \special{pa 3132 900}%
    \special{pa 3559 1488}%
    \special{da 0.050}%
    \special{pa 3559 1488}%
    \special{pa 4250 1263}%
    \special{da 0.050}%
    \special{pa 4250 1263}%
    \special{pa 4250 537}%
    \special{da 0.050}%
    \graphtemp=.5ex
    \advance\graphtemp by 0.900in
    \rlap{\kern 1.650in\lower\graphtemp\hbox to 0pt{\hss $\R$\hss}}%
    \graphtemp=.5ex
    \advance\graphtemp by 0.000in
    \rlap{\kern 0.750in\lower\graphtemp\hbox to 0pt{\hss $i\R$\hss}}%
    \graphtemp=.5ex
    \advance\graphtemp by 1.900in
    \rlap{\kern 0.750in\lower\graphtemp\hbox to 0pt{\hss Even example $d=4$\hss}}%
    \graphtemp=.5ex
    \advance\graphtemp by 0.900in
    \rlap{\kern 4.650in\lower\graphtemp\hbox to 0pt{\hss $\R$\hss}}%
    \graphtemp=.5ex
    \advance\graphtemp by 0.000in
    \rlap{\kern 3.750in\lower\graphtemp\hbox to 0pt{\hss $i\R$\hss}}%
    \graphtemp=.5ex
    \advance\graphtemp by 1.900in
    \rlap{\kern 3.750in\lower\graphtemp\hbox to 0pt{\hss Odd example $d=5$\hss}}%
    \hbox{\vrule depth1.900in width0pt height 0pt}%
    \kern 4.650in
  }%
}%

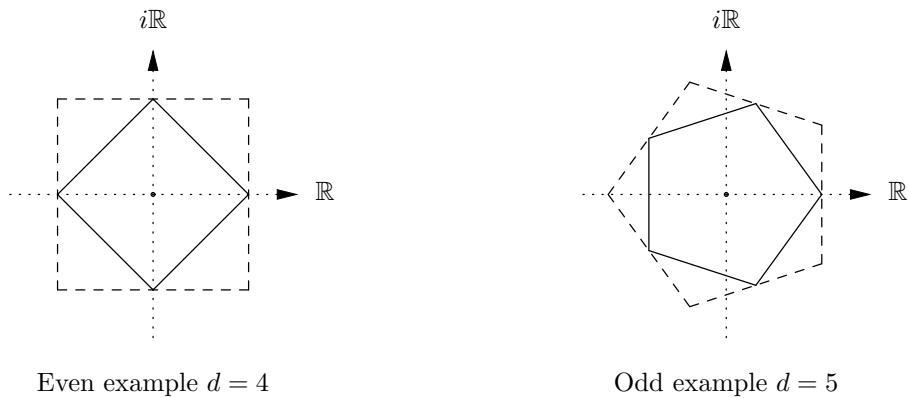
\figure Figure. {The boundaries of the convex hull of~$\U$ (solid) and its dual (dashed)}
       {$$\box\graph$$}
\endinsert

\label{HullUinequality}
\stat Lemma.  Define $\rho=\exp(i\pi/d)$.  For each $\beta\in\Hull\U$, the following inequality
holds:
$$
  \re(\rho\beta) \leq \cos(\pi/d).
$$

\proof From Lemma~\ref{Ucirc}, we have $\U^\circ={\rho\over\cos(\pi/d)}\Hull\U$.  Hence 
$\rho\beta=\cos(\pi/d)\gamma$ for some $\gamma\in\U^\circ$.  Thus, 
$\re(\rho\beta)=\cos(\pi/d)\re(\gamma$).  But we have $\re(\gamma)=\re\pairing1\gamma\leq1$ 
because $1\in\U$.  The result follows because $\cos(\pi/d)>0$ (we assumed 
$d>2$, note also that case $d=2$ is trivially true).  \qed

\subsection Duality and DFT

  As in Section~\secref{MDFT}, we put $D=d^n$.  The map DFT is linear and its matrix 
$U=H_d^{\circ n}$ (in the canonical basis of~$\C^D$) satisfies $U^\dagger U=DI$, where $U^\dagger$
is the conjugate transpose of~$U$.  This has some useful consequences.

\label{pairinghat}
\stat Lemma.  Assume that $\beta,\gamma\in\C^D$, and put $\hat\beta=\DFT\beta$ and 
$\hat\gamma=\DFT\gamma$.  We have $\pairing{\hat\beta}{\hat\gamma}=D\pairing\beta\gamma$.

\proof If we identify $\beta$ and $\gamma$ with the column vectors of their coordinates in the 
canonical basis we can write:
$$
  \pairing{\hat\beta}{\hat\gamma}
  =
  \hat\beta^\dagger \hat\gamma
  =
  (U\beta)^\dagger U\gamma
  =
  \beta^\dagger U^\dagger U \gamma
  =
  D \beta^\dagger \gamma
  =
  D \pairing\beta\gamma
$$
as claimed.  \qed

\label{DFThat}
\stat Proposition.  Let $\Gamma$ be a polytope in~$\C^D$ containing~0, and denote by~$\hat\Gamma$
its image (which is also a polytope, by linearity of~DFT) under the map~DFT.  We have the following
relations between their duals:
$$
  \widehat{\Gamma^\circ}
  =
  D\,\hat\Gamma^\circ.
$$

\proof For $\beta\in\C^D$, we have $\beta\in\Gamma^\circ$ if and only if
$\pairing\beta\gamma\leq1$ for all $\gamma\in\Gamma$.  From Lemma~\ref{pairinghat}, this is 
equivalent to $\pairing{\hat\beta}{\hat\gamma}\leq D$ for all $\gamma\in\Gamma$.  This condition 
can be written $\pairing{{1\over D}\hat\beta}{\hat\gamma}\leq1$, or therefore
${1\over D}\hat\beta\in\hat\Gamma^\circ$.
Finally, it is equivalent to $\hat\beta\in D\,\hat\Gamma^\circ$.  \qed

\label{HBineq}
\section Homogeneous Bell inequalities

  Le $n$ be the number of parties.  For each party, we consider two observables, denoted by 
$\hat A_i$ and~$\hat B_i$ (for $1\leq i\leq n$).  The outcomes of each measure are assumed to belong
to the set~$\U$ defined in~\eqref{setU}, with $d\geq2$.

Recall, from the identity
$$
  1 - X^d
  =
  (1 - X) (1 + X + X^2 + \cdots + X^{d-1}),
$$
that the roots of the polynomial $1+X+\cdots+X^{d-1}$ are the elements of $\U\setminus\{1\}$.
Recall also that $\sum_{u\in\U}u^k$ evaluates to $d$ when $k$ is a multiple of $d$ but is zero
otherwise.  If $a_i,b_i\in\U$, there exists an integer $r_i\in\Z_d$ such that 
$a_i/b_i=\omega^{r_i}$.  Let also $s_i\in\Z_d$.  Then
$$
  \eqalign{  
       a_i^{d-1} + \omega^{s_i} a_i^{d-2}b_i + \cdots + \omega^{(d-1)s_i} b_i^{d-1}
    &= a_i^{d-1} (1 + \omega^{s_i-r_i} + \cdots + \omega^{(d-1)(s_i-r_i)}) \cr
    &= \left\{\eqalign{
           a_i^{d-1}d &\qquad\hbox{if $r_i=s_i$,} \cr
           0         &\qquad\hbox{otherwise.}\cr}
       \right.\cr}
$$
Let now $f$ be any map from $\Z_d^n$ to $\U$.  We have
$$
  \sum_{s\in\Z_d^n} f(s) 
     \prod_{i=1}^n (a_i^{d-1} + \omega^{s_i} a_i^{d-2}b_i + \cdots +
           \omega^{r_is_i} a_i^{d-1-r_i}b_i^{r_i} + \cdots + \omega^{(d-1)s_i} b_i^{d-1})
  =
  u d^n
  \eqdef{sumstodn}
$$
where $u\in\U$, because in this sum, exactly one term is non-zero (the one corresponding to
$s_i=r_i$ for each~$i$).

  If we expand the products in~\eqref{sumstodn}, we get
$$
  \eqalign{ud^n
   &= \sum_{s\in\Z_d^n} f(s) 
      \sum_{r\in\Z_d^n} \prod_{i=1}^n \omega^{s_ir_i} \, a_i^{d-1-r_i}b_i^{r_i} \cr
   &= \sum_{s\in\Z_d^n} f(s) \sum_{r\in\Z_d^n} \omega^{r\cdot s} \, a^r
      \qquad\hbox{where }a^r\assign\prod_{i=1}^n a_i^{d-1-r_i}b_i^{r_i} \cr
   &= \sum_{r\in\Z_d^n} \hat f(r)\,a^r \qquad\hbox{where $\hat f=\DFT f$.}\cr}
$$
Now, if the $a_i$ and~$b_i$ are random variables we can write, about expected values:
$$
  \sum_{r\in\Z_d^n} \hat f(r)\,E(a^r) \in d^n\,\Hull\U.
$$
From Lemma~\ref{HullUinequality}, we obtain:
$$
  \re\bigg( \rho \sum_{r\in\Z_d^n} \hat f(r)\,E(a^r) \bigg)
  \leq 
  d^n \cos(\pi/d).
$$
When $d>2$, this also can be written:
$$
  \re\bigg(
      {\rho\over d^n\cos(\pi/d)}  \sum_{r\in\Z_d^n} \hat f(r) E(a^r) 
  \bigg) 
  \leq 1
  \qquad\hbox{for $f\in\F_{d,n}$.}
  \eqdef{ditBell}
$$
We call these relations homogeneous Bell inequalities.  There are $d^D$ of them.

\label{Polynomials}
\section Homogeneous Bell polynomials 

  We now study the polynomials in $2n$ variables $A_i$ and $B_i$ (for $1\leq i\leq n$) which are
involved in the homogeneous Bell inequalities.  Some Bell polynomials where defined 
in~\bib{WernerWolf} for $d=2$.  As a generalisation to the multidimensional case, we define the 
homogeneous Bell polynomials to be
$$
  \P_f
  =
  \sum_{r\in\Z_d^n} \hat f(r) A^r
  \qquad\hbox{where }A^r\assign\prod_{i=1}^n A_i^{d-1-r_i}B_i^{r_i}
  \eqdef{HBP}
$$
where $f$ is any map from $\Z_d^n$ to~$\U$.  Let us denote $\H_{d,n}$ the set of these polynomials.
Each element of~$\H_{d,n}$ is a homogeneous polynomial of degree $n(d-1)$.  Note that in view of
Section~\secref{violations}, we consider $\P_f$ as a {\it non commutative} polynomial.  More
precisely, each $A_i$ is {\bf not} assumed to commute with~$B_i$, while $A_i$ and $B_i$ do commute
with $A_j$ and~$B_j$ for $i\neq j$.

  As in~\bib{Schaefer} where the case $d=2$ is handled, we give a recursive construction of the
homogeneous Bell polynomials.  This construction is a direct consequence of
Equation~\eqref{Hrecursion}.  If $\P_0,\ldots,\P_{d-1}$ are homogeneous Bell polynomials
in the~$2(n-1)$ the variables $A_i,B_i$ with $1\leq i\leq n-1$, then we get a homogeneous Bell
polynomial in $2n$ variables by the $d$-ary operation~$\bowtie$:
$$
  \P_0 \bowtie \cdots \bowtie \P_{d-1}
  \assign
  \sum_{r_n=0}^{d-1}  
     \bigg( \sum_{t=0}^{d-1} \omega^{r_nt} \P_t  \bigg)
     A_n^{d-1-r_n}B_n^{r_n}.
$$
Conversely, every element of the set~$\H_{d,n}$ can be obtained this way.

  For example, with $d=2$, the polynomials obtained are $\pm1$ for $n=0$, $\pm2 A_1$ and $\pm2B_1$
for $n=1$, and
$$
  \eqalign{
    &\pm4A_1A_2, \cr
    &\pm4A_1B_2, \cr
    &\pm4B_1A_2, \cr
    &\pm4B_1B_2, \cr}
  \qquad
  \eqalign{
      \pm2(-A_1A_2+A_1B_2+B_1A_2+B_1B_2)&, \cr
      \pm2(A_1A_2-A_1B_2+B_1A_2+B_1B_2)&, \cr
      \pm2(A_1A_2+A_1B_2-B_1A_2+B_1B_2)&, \cr
      \pm2(A_1A_2+A_1B_2+B_1A_2-B_1B_2)&, \cr}
$$
for $n=2$ (we recognize the polynomials involved in the CHSH inequalities).  Examples for $d=3$ 
will be given in Section~\secref{dthree}.

\subsection Symmetries

  The set~$\H_{d,n}$ of homogeneous Bell polynomials has some symmetries we briefly discuss now.
They are consequences of Proposition~\ref{easyresults}.

{\bf a.}  If the maps $f$ and $g\in\Z_d^n$ are the same, up to the order of their
arguments:
$$
  g(s_1,\ldots,s_n)
  =
  f(s_{\sigma(1)},\ldots,s_{\sigma(n)})
  \qquad\hbox{for all $s\in\Z_d^n$,}
$$
for some permutation~$\sigma$, then the polynomial $\P_g$ can be obtained from~$\P_f$ by changing
each variable $A_i$ (resp.~$B_i$) to $A_{\sigma(i)}$ (resp.~$B_{\sigma(i)}$).  This
symmetry corresponds to the fact that the $n$ subsystems are indistinguishable.  

{\bf b.} If, for some $i_0$,
$$
  g(s)
  =
  \omega^{-s_{i_0}} f(s)
  \qquad\hbox{for all $s\in\Z_d^n$,}
$$
then, from Proposition~\ref{easyresults}, we have $\hat g(r)=\hat f(r-\delta)$ for all $r\in\Z_d^n$,
where $\delta=(0,\ldots,0,1,0,\ldots,0)$ has its only non-null component at index~$i_0$.  Hence, we 
obtain 
$$
  \P_g = \sum_{r\in\Z_d^n} \hat g(r) A^r
  = \sum_{r\in\Z_d^n} \hat f(r-\delta) A^r
  = \sum_{r\in\Z_d^n} \hat f(r) A^{r+\delta}.
$$
This shows that we obtain $\P_g$ from $\P_f$ by the circular monomial substitution 
$$
  A_{i_0}^{d-1}
  \longrightarrow
  A_{i_0}^{d-2}B_{i_0}
  \longrightarrow
  \cdots
  \longrightarrow
  A_{i_0}B_{i_0}^{d-2}
  \longrightarrow
  B_{i_0}^{d-1}
  \longrightarrow
  A_{i_0}^{d-1}.
$$
Also, the set~$\H_{d,n}$ is invariant, under the swap operation $A_{i_0}\leftrightarrow B_{i_0}$
(this can be algebraically checked with the help of Proposition~\ref{easyresults}(a)).
Hence, for each $i_0$, the set $\H_{d,n}$ is invariant under the action of the dihedral 
group of order~$2d$ over the monomials made of the variables $A_{i_0}$ and~$B_{i_0}$.

{\bf c.}  Of course, the set $\H_{d,n}$ is also invariant under multiplication by $\omega$,
and by complex conjugation (Proposition~\ref{easyresults}(b) can be used to check this latter fact).

\section The classical domain

  We now show that the homogeneous Bell inequalities obtained in Section~\secref{HBineq} are tight 
and completely characterize a local realistic model, for $n\in\N^*$ parties, $m=2$ observables for 
each site, and $d$-outcomes measurements with $d>2$. 

  The values $a_i$ and~$b_i$, when a local realistic model is applied, of these $2n$ observables are
assumed to belong to the set $\U$.   We consider the monomials 
$$
  A^s
  =
  \prod_{i=1}^n A_i^{d-1-s_i}B_i^{s_i}
$$
which appear in homogeneous Bell polynomials.  There are $D=d^n$ of them.  
For each experiment, the data set of the values obtained for these monomials form a vector
$\xi=(a^s)_{s\in\Z_d^n}$ in $\C^D$.  Our aim is to show that the domain accessible to the expected 
values of~$\xi$ is the polytope defined by the inequalities~\eqref{ditBell}.

\subsection The polytope $\Omega$

  Put
$$
  \xi_r
  =
  (\omega^{r\cdot s})_{s\in\Z_d^n}
  \in\C^D
  \qquad\hbox{for each $r\in\Z_d^n$.}
$$
The $d^{n+1}=dD$ vectors $u\xi_r$, for $u\in\U$ and $r\in\Z_d^n$ are all distinct.
In a local realistic model, each experimental data set assigns a value
$$
  \prod_{i=1}^n a_i^{d-1-s_i}b_i^{s_i}
  =
  \prod_{i=1}^n a_i^{d-1} \prod_{i=1}^n \omega^{r_is_i}
  =
  \prod_{i=1}^n a_i^{d-1} \omega^{r\cdot s}
$$
to each monomial $A^s$ where $\omega^{r_i}=b_i/a_i$ (for $1\leq i\leq n$).  Thus, the vector~$\xi$
obtained from experimental data is one of the vectors $u\xi_r$, where $u=\prod_{i=1}^na_i^{d-1}$, and
$r=(r_i)_{1\leq i\leq n}$ with the $r_i$ just defined.

  Conversely, it is possible to design classical experiments which assign independently any value
in~$\U$ to the $2n$ variables and which assign any $u\xi_r$ to the data set vector~$\xi$.
Then, if the values assigned to the variables follow some probability distributions, expected
values for the vectors~$\xi$ obtained, are convex combinations of the $u\xi_r$.  Hence the
classically accessible region for $\xi$ is the convex hull of the $u\xi_r$, which will be denoted
by~$\Omega$ as it was in~\bib{WernerWolf} for the case $d=2$.  The domain~$\Omega$ is a polytope
in~$\C^D$ and has $dD$ vertices.  Notice that $\Omega$ has a $d$-order symmetry:
$\omega\Omega=\Omega$.

\subsection The polytope $\Pi=\DFT^{-1}\Omega$

   We can find all the inequalities defining the facets of the polytope~$\Omega$.  They will be
the $d^D$ homogeneous Bell inequalities~\eqref{ditBell} we obtained in Section~\secref{HBineq}.

  Let $(\pi_s)_{s\in\Z_d^n}$ be the canonical basis of the complex vector space~$\C^D$.  The
discrete Fourier transform maps the $\pi_s$ to the $\xi_s$.  We consider the following polytope:
$$
  \Pi
  \assign
  \Hull \{u\pi_s \mid u\in\U, s\in\Z_d^n\}.
$$
Then $\Omega$ is $\hat\Pi$, the image of~$\Pi$ under DFT.
To find the facets of $\Omega$, we have to study its dual.  But from Proposition~\ref{DFThat},
$$
  \Omega^\circ = \hat\Pi^\circ = {1\over d^n}\,\widehat{\Pi^\circ}.
  \eqdef{Omegacirc}
$$
Let's first study~$\Pi^\circ$.

\label{Picirc}
\stat Proposition.  The vertices of the polytope $\Pi^\circ$ are the $\beta=(\beta_1,\ldots,\beta_s)$
such that
$$
  \beta_s 
  =
  {\rho\over\cos(\pi/d)} f(s)
  \qquad\hbox{where $f$ is any element of~$\F_{d,n}$.}
$$

\proof By definition,
$$
  \Pi^\circ
  =
  \{\beta\in\C^D \mid \re\pairing\beta{u\pi_s} \leq 1, \forall u\in\U, s\in\Z_d^n\}.
$$
Using the $d$-order symmetry of~$\U^\circ$, and using 
$\pairing\beta{u\pi_s}=u\pairing\beta{\pi_s}$, we can write
$$
  \Pi^\circ
  =
  \{\beta\in\C^D \mid \pairing\beta{\pi_s} \in \U^\circ, \forall s\in\Z_d^n\}.
$$
We are interested with the extremal points of~$\Pi^\circ$.  These are obtained when
$\pairing\beta{\pi_s}$ are in a corner of~$\U^\circ$ (see Lemma~\ref{Ucirc}):
$$
  \pairing\beta{\pi_s}
  \in
  \V
  =
  {\rho\over\cos(\pi/d)}\U.
$$
Hence, there exists $f\in\F_{d,n}$ such that:
$$
  \beta_s^* = \pairing\beta{\pi_s} 
  =
  {\rho\over\cos(\pi/d)}f(s)
  \qquad\hbox{for all $s\in\Z_d^n$.}
$$
But $\Pi^\circ$ is symmetric under complex conjugation.  Hence we can change $\beta_s^*$ 
for~$\beta_s$. \qed

\subsection The dual of $\Omega$

\stat Theorem.  The vertices of the polytope $\Omega^\circ$ are given by
$$
  {\rho\over d^n\cos(\pi/d)} \big(\hat f(r)\big)_{r\in\Z_d^n}
  \qquad\hbox{for $f\in\F_{d,n}$.}
$$

\proof The result follows from Equation~\eqref{Omegacirc} and Proposition~\ref{Picirc}.  \qed

  To end this section, note that the inequalities~\eqref{ditBell} can be written
$$
  \re\pairing{\beta_f}\xi \leq 1
  \qquad\hbox{with}\quad
  \beta_f^* = {\rho\over d^n\cos(\pi/d)} \big(\hat f(r)\big)_{r\in\Z_d^n} 
  \quad\hbox{and}\quad
  \xi = \big(E(a^r)\big)_{r\in\Z_d^n}.
$$
Hence the theorem just obtained shows that our homogeneous Bell inequalities define the facets
of the polytope~$\Omega$.  Thus they form a complete set of tight Bell inequalities.

\label{violations}
\section Violations by Quantum Mechanics

  At this point, we have only considered local-realistic models.  The polytope~$\Omega$ we have
made explicit using homogeneous Bell inequalities is the domain accessible with such models.
However, the primary aim of Bell inequalities was (at least historically) to compare 
local-realistic theories with Quantum Mechanics.  The main success of the original and CHSH Bell
inequalities, was due to the fact that Quantum Mechanics violate them, hence they provided the
proof that quantum indeterminacy cannot be explained by hidden variables.  We now show
that Quantum Mechanics also violates homogeneous Bell inequalities, and that this fact could be, in
principle, checked by experiment.

  There exists a difficulty in our setting, which did not appear in the $d=2$ case.  
Multidimensional homogeneous Bell polynomials involve products of variables, some of them 
corresponding to observables of the same party.  In Quantum Mechanics, such observables corresponds
to non commuting operators and their values cannot be simultaneously obtained, and this prevents to
observe violations this way.  However, there are important cases where such products of observables
are themselves observables.  This is our key tool now.

\subsection Generalized Pauli matrices

  We use the following multidimensional generalization (found for example in~\bib{Schwinger}
and~\bib{Gottesman}) of Pauli (or spin) matrices.  Let
$$
  X 
  = 
  \pmatrix{
     0 & 0 & \cdots & 1 \cr
     1 & \ddots  & \ddots & \vdots \cr
     \vdots & \ddots & \ddots & 0 \cr
     0 & \cdots & 1 & 0 \cr}
  =
  \sum_{i=0}^{d-1} \ket{i+1\bmod d}\bra i
  \qquad\hbox{and}\qquad
  Z
  =
  \pmatrix{
    1 & 0 & \cdots & 0 \cr
    0 & \omega & \ddots & 0 \cr
    \vdots & \ddots & \ddots & \vdots \cr
    0 & 0 & \cdots & \omega^{d-1} \cr}
  =
  \sum_{i=0}^{d-1} \omega^i\ket i\bra i,
$$
where the kets~$\ket0,\ldots,\ket{d-1}$ form an orthonormal basis of~$\C^d$ (in fact, the
eigenbasis of~$Z$).  The matrices $X$ and~$Z$ have order~$d$ and satisfy $ZX=\omega XZ$.  
The {\it generalized Pauli matrices} are the following $d+1$ unitary matrices:
$$
  Z,\ X,\ XZ,\ \ldots,\ XZ^{d-1}.
  \eqdef{genPauli}
$$
The following two results are easy to show.  The first one is about eigenvalues as these are
the possible outcomes of measurements in Quantum Mechanics.

\stat Proposition.  Let $k$ be an integer.  The eigenvalues of $XZ^k$ are the $\omega^j$ (with
$0\leq j\leq d-1$) when $d$ is odd or when $k$ is even.  They are the $\rho\omega^j$, with
$\rho=\exp(i\pi/d)$, when $d$ is even and $k$ is odd. 

\proof  By expanding the characteristic polynomial of $XZ^k$ along the last column, we obtain~:
$$
  \eqalign{\det{(XZ^k - \lambda I)}
    &= (-1)^{d-1} \omega^{k(d-1)} \omega^{k(0+1+\cdots+(d-2))}-\lambda(-\lambda^{d-1}) \cr
    &= (-\lambda)^d - (-1)^d \omega^{k(d-1)+k(d-1)(d-2)/2}  
     = (-\lambda)^d - (-1)^d \omega^{k(d-1)d/2} \cr
    &= \left\{\eqalign{
           -\lambda^d + 1^{k(d-1)/2} = 1 - \lambda^d &\quad\hbox{when $d$ is odd,}\cr
            \lambda^d - (-1)^{k(d-1)} = \lambda^d - (-1)^k &\quad\hbox{when $d$ is even.}\cr} 
      \right.\cr}
$$
Hence the eigenvalues are the solutions of equation $\lambda^d=1$ when $d$ is odd or $k$ is even~;
and of $\lambda^d=-1$ when $d$ is even and $k$ is odd.  \qed

\label{Paulipower}
\stat Lemma.  For any integers $k,e$, the following relation holds
$$
  (XZ^k)^e 
  =
  \omega^{ke(e-1)/2} X^eZ^{ke}.
$$

\proof  We leave it to the reader.  It can be done by induction over~$e$, using 
$ZX=\omega XZ$. \qed

\subsection Unitary observables

It is shown in~\bib{BBRV}, and also in~\bib{PittengerRubin}, that when $d$ is a power of a prime,
the bases consisting of the normalized eigenvectors of the $d+1$ Pauli matrices given
by~\eqref{genPauli} form $d+1$ {\sl Mutually Unbiased Bases}~\bib{Ivanovic}.  Paterek~\bib{Paterek}
explains that these generalized Pauli matrices can be used as unitary observables,
instead of more classical Hermitian operators. (Note that for the usual case $d=2$,
the matrices $X$ and~$Z$ are Hermitian as well as unitary.)  These are clearly the operators we
need, as we considered complex valued observables.

  To determine quantum violations of homogeneous Bell inequalities, we have to evaluate expected
values of operators of the form $X^{d-1-r}Z^r$, for $0\leq r\leq d-1$.  These are unitary
observables, and it should be possible to directly obtain outcomes of them (without
measuring outcomes of~$X$ and~$Z$).  At least, we can rely on the better known generalized Pauli
operators using the following Proposition.

\stat Proposition.  Let $0\leq r\leq d-1$ and assume $d$ prime.  It is possible to experimentally 
obtain values for $X^{d-1-r}Z^r$ in order to compute the corresponding expected value.

\proof When $r=d-1$, just make a measurement with operator $Z$ on each sample, and raise the
outcomes to power~$d-1$.  We now assume that $0\leq r\leq d-2$.  Thus $1\leq d-1-r\leq d-1$.  As
$d$ is prime, then $(d-1-r)$ is invertible modulo~$d$ and it is possible to find an integer $k$
such that $1\leq k\leq d-1$ and $k(d-1-r)\equiv r$ modulo~$d$.  From Lemma~\ref{Paulipower},
we get
$$
  (XZ^k)^{d-1-r}
  =
  \omega^{k(d-1-r)(d-2-r)/2} X^{d-1-r} Z^{k(d-1-r)}
  =
  \omega^{k(r+1)(r+2)/2} X^{d-1-r}Z^r.
$$
Hence, we have to make a measurement with operator $XZ^k$ on each sample, raise the outcomes to
the power $d-1-r$, and multiply the results with $\omega^{-k(r+1)(r+2)/2}$.   \qed

   Now, we are able to compute some violations of homogeneous Bell inequalities
by Quantum Mechanics, with the quantum operators $X$ and~$Z$ in place of the classical operators
$A_i$ and~$B_i$ respectively.  Hence, we consider the following quantum counterparts of our 
homogeneous Bell polynomials~\eqref{HBP}:
$$
  \Q_f
  =
  \sum_{r\in\Z_d^n} \hat f(r) A_{QM}^r
  \qquad\hbox{where }A_{QM}^r\assign\bigotimes_{i=1}^n (X^{d-1-r_i}Z^{r_i}).
$$
A quantum state $\ket\phi$ will violate the corresponding homogeneous Bell inequality if
the condition
$$
  \re\Big ({\rho\over d^n\cos(\pi/d)} \langle\psi|Q_f|\psi\rangle \Big)
  \leq 1
$$
is {\bf not} satisfied.  Our short study of the case $d=3$ will indeed exhibit cases where
such violations occur.

\label{dthree}
\section The case $d=3$

  We illustrate our results with the first multidimensional case: $d=3$ (sometimes called
trichotomic).  Note that the factor~$1/\cos(\pi/d)$ in Equation~\eqref{ditBell} is maximal in this
case, and this might lead to higher violations.

\subsection DFT

  Here, $\omega=\exp(2i\pi/3)$ and
$$
  H_3^{\otimes1} = H_3
  = \pmatrix{
       1 & 1 & 1 \cr
       1 & \omega & \omega^2 \cr
       1 & \omega^2 & \omega \cr}
  \qquad\qquad
  H_3^{\otimes2} = \pmatrix{
    1 & 1 & 1 & 1 & 1 & 1 & 1 & 1 & 1 \cr
    1 &\omega   &\omega^2 &1 &\omega   &\omega^2 &1 &\omega   &\omega^2 \cr
    1 &\omega^2 &\omega   &1 &\omega^2 &\omega   &1 &\omega^2 &\omega   \cr
    1 &1        &1        &\omega &\omega   &\omega   &\omega^2 &\omega^2 &\omega^2 \cr
    1 &\omega   &\omega^2 &\omega &\omega^2 &1        &\omega^2 &1        &\omega   \cr
    1 &\omega^2 &\omega   &\omega &1        &\omega^2 &\omega^2 &\omega   & 1       \cr
    1 &1        &1        &\omega^2 &\omega^2 &\omega^2 &\omega &\omega   &\omega   \cr
    1 &\omega   &\omega^2 &\omega^2 &1        &\omega   &\omega &\omega^2 &1        \cr
    1 &\omega^2 &\omega   &\omega^2 &\omega   &1        &\omega &1        &\omega^2 \cr}.
$$

\subsection The hull of $\U$ and its dual

The hull of~$\U$ is the triangle with vertices 1, $\omega$, $\omega^2$ and its edges are defined 
by the three inequalities
$$
  x + \sqrt3y \leq1,
  \qquad
  -2x \leq 1,
  \qquad
  x - \sqrt3y \leq 1.
$$
Hence, the dual~$\U^\circ$ has vertices $1+i\sqrt3, -2, 1-i\sqrt3$, which are obtained from the
vertices of $\Hull\U$ by multiplication by~$\exp(i\pi/3)/\cos(\pi/3)=-2\omega^2$.

\subsection Bell polynomials

  We did some computations, with the help of the \|Magma| computer algebra system.  For the
(virtual) case $n=0$, the trichotomic Bell polynomials are the constant polynomials 1, $\omega$
and~$\omega^2$.   For $n=1$, there are yet 27 homogeneous trichotomic Bell polynomials.  Instead of
listing them all, we give for them the following compact expression:
$$
  u \big(3M + (v - 1)(A^2 + AB + B^2)\big)
  \eqdef{weirdformula}
$$
where $u,v\in\{1,\omega,\omega^2\}$ and $M\in\{A^2,AB,B^2\}$.

  For $n=2$, there are 19683 homogeneous trichotomic Bell polynomials.  Among them, 18792 are
irreducible polynomials.  The number of elements in~$\H_{3,2}$ with only real coefficients is~81 
(the aim of this criterion here is just to reduce the list size).  We can list them, up to the 
symmetries discussed in Section~\secref{Polynomials}, as there remain only 4 ones:
$$
  \displaylines{
     9A_1^2A_2^2 \cr
     3(A_1^2A_2^2 - A_1^2B_2^2 + 2A_1B_1A_2B_2 + A_1B_1B_2^2 - B_1^2A_2^2 + B_1^2A_2B_2) \cr
     3(-A_1^2A_2B_2 + A_1^2B_2^2 + A_1B_1A_2^2 + 2A_1B_1B_2^2 - B_1^2A_2^2 + B_1^2A_2B_2) \cr
     3(2A_1^2A_2^2 - A_1^2A_2B_2 - A_1^2B_2^2 + A_1B_1A_2^2 + A_1B_1A_2B_2 + A_1B_1B_2^2). \cr}
$$
We found also that there are 243 elements in~$\H_{3,2}$ up to these symmetries.

\subsection Bell inequalities

  The factor ${\rho\over d^n\cos(\pi/d)}$ with appear in Inequalities~\eqref{ditBell} is in this
case $-2\omega^2/3^n$.  By changing $f$ to $\omega f$, we can remove the $\omega^2$ to obtain
the following homogeneous trichotomic Bell inequalities:
$$
  -\re\bigg(
      {2\over 3^n}  \sum_{r\in\Z_3^n} \hat f(r) E(a^r) 
  \bigg) 
  \leq 1
  \qquad\hbox{for each $f\in\F_{3,n}$.}
  \eqdef{triBell}
$$

\subsection Violations 

  Yet the case $n=1$ is especially interesting.  The 27 homogeneous Bell polynomials fall in 3
classes according to the value of~$v$ in formula~\eqref{weirdformula}.  The most interesting class
is the one obtained with $v=\omega^2$.  In that case, the eigenvalues of the operator obtained
are $-3\zeta$, $-3\zeta\omega$ and $-3\zeta\omega^2$ where $\zeta=\exp(2i\pi/9)$.  
They do not belong to $\Hull\U$.  In particular $-2\re(-3\zeta)/3\simeq1.53209>1$ and one can
expect violations. This is indeed the case: consider the map $f$ such that $f(0)=\omega$ and
$f(1)=f(2)=\omega^2$.  Then we have $\hat f(0)=\omega^2-1$, $\hat f(1)=\hat f(2)=\omega-\omega^2$
and Equation~\eqref{triBell} reads
$$
  -{2\over3} 
  \re\Big((\omega^2-1)E(a^2) + (\omega-\omega^2) \big(E(ab) + E(b^2)\big)\Big)
  \leq 1.
$$
The corresponding operator is
$$
  \Q_f
  =
  (\omega^2-1) X^2 + (\omega-\omega^2) (XZ + Z^2)
  =
  \pmatrix{ 
     \omega-\omega^2 & \omega^2 - 1 & 1 - \omega   \cr
     \omega-\omega^2 & 1 - \omega   & \omega^2 - 1 \cr
     \omega^2 - 1    & \omega^2 - 1 & \omega^2 - 1 \cr}
$$
and has $\lambda=-3\zeta$ as an eigenvalue.   Then we can find states~$\ket\psi$ such that
$-2\re\langle\psi|Q_f|\psi\rangle/3$ exceeds~1.  For example, the state
$(\ket0+2\ket1+3\ket2)/\sqrt{14}$ achieves a violation of~$19/14\simeq1.357$ and the state 
$$
  \big( (44+50\omega)\ket0 + (76+9\omega)\ket1 + (143+17\omega)\ket2\big) / \sqrt{25716}
$$
achieves a violation of~$1.53208$.  Non-locality is not needed to violate homogeneous Bell
inequalities!  Of course, this situation did not appear in dimension $d=2$, as there were only
trivial Bell polynomials with $n=1$.
  
  For $n=2$, the best violation is obtained in 27 cases for which eigenvalues are $9(1-\omega)$,
$9(\omega^2-1)$, $9(\omega-\omega^2)$ and 0 (with multiplicity~6).  One of these cases is the
following
$$
  \Q_f 
  =
  3\Big( (\omega^2-1)X^2\otimes XZ + (\omega^2-1)XZ\otimes X^2 + (1-\omega)Z^2\otimes Z^2 \Big)
$$
obtained from~\eqref{triBell} with the map $f$ whose vector of values is 
$(\omega^2,\omega,\omega^2,\omega,\omega,1,\omega^2,1,1)$.
We expect violations of $-2\re\big(9(\omega^2-1)\big)/9=3$.
Such violation is obtained with the state $(\ket{01}+\ket{10}+\omega\ket{22})/\sqrt3$.

\section Conclusion

  In this paper, we defined homogeneous Bell inequalities and we showed that they correspond to 
the boundaries of the domain accessible with local-realistic models, for the general multipartite 
and multidimensional case with two observables per party.  
We studied homogeneous Bell polynomials and their symmetries.  
It turns out that the classical domain is the image under DFT of a polytope obtained 
from the canonical basis, and we used this fact to compute its dual.  With this, we were able to
show that the homogeneous Bell inequalities form a complete set.

  Then we considered violations by Quantum Mechanics, using the observables provided by generalized
Pauli matrices.  We showed that violations indeed occur, and exhibit some of them in the
trichotomic case.

  The complex valued correlation function we used is a natural mathematical generalisation of the 
two-dimensional one.  Fu in~\bib{Fu} argued that it has also a physical meaning, at least in 
Quantum Mechanics.  It was a crucial and fruitful ingredient in the present work, and this raises
interrogations about the precise extent of this physical meaning.  Also, complex valued observables
provided by the generalized Pauli matrices were a key tool for computing violations.

\section Bibliography

\showbib

\bye